\documentclass[%
aip,
jcp,
preprint, 
amsmath,amssymb,
]{revtex4-2}

\usepackage{graphicx}
\usepackage{dcolumn}
\usepackage{bm}

\usepackage[utf8]{inputenc}
\usepackage[T1]{fontenc}
\usepackage{mathptmx}
\usepackage{etoolbox}
\usepackage[version=4]{mhchem}
\usepackage{soul}
\usepackage{siunitx}

\makeatletter
\def\@email#1#2{%
	\endgroup
	\patchcmd{\titleblock@produce}
	{\frontmatter@RRAPformat}
	{\frontmatter@RRAPformat{\produce@RRAP{*#1\href{mailto:#2}{#2}}}\frontmatter@RRAPformat}
	{}{}
}%
\makeatother

\begin{document}
	
	
	\title[Beyond the Static Kuhn Length]{Beyond the Static Kuhn Length: Conformational Substructures and Relaxation Dynamics in Flexible Chains}
	\author{José  A. Martins}
	\affiliation{Departamento de Engenharia de Polímeros, Universidade do Minho, Campus de Azurém, 4800-058 Guimarães, Portugal} 
	\email{jamartins@dep.uminho.pt}
\noindent\href{https://orcid.org/0000-0003-1164-8411}{ORCID: 0000-0003-1164-8411}

	\date{\today}%

%

\begin{abstract}
The statistical, ``monomer-based'' segment length $b$ and the Kuhn length $l_k$ are central to polymer physics, yet the minimal size required for truly statistical segment—Gaussian, uncorrelated, and valid as an entropic spring—is not rigorously established. Using atomistic simulations of entangled polyethylene, we re-evaluate these foundational quantities.

By fitting end-to-end distance distributions of C–C bond blocks to a Gaussian form and validating with higher-moment analyses, we identify for the first time the minimal sizes corresponding to a statistical segment and an entropic spring. A single Kuhn segment ($\approx$11 bonds) is the smallest statistically uncorrelated unit, but its distance distribution is strongly non-Gaussian, while the ``monomer-based'' segment $b$, used in rheology and classical tube-theory formulations, is not statistical at all.  True Gaussianity emerges only for blocks containing multiple Kuhn segments.

At the Kuhn scale, we uncover a previously unresolved heterogeneity. Each segment samples a broad conformational range, from coiled ($\approx 4$~\AA) to extended ($\approx 14$~\AA), giving rise to distinct types of Kuhn segments. These organize into three categories: aligned chain segments (ACS), random conformational sequences (RCS), and chain ends (CE). Each type exhibits its own dynamical signature. ACS relax with $\beta \approx 0.5$, consistent with quasi-one-dimensional, defect-mediated localized modes, whereas RCS and CE relax with $\beta \approx 0.7$. Translational motion is likewise heterogeneous: all segment types exhibit subdiffusion $g_1(t)\propto t^{0.7}$ within the Kuhn-scale dynamical window.

By connecting these observations to the localized-mode theory of Skolnick-Helfand and the continuous-time random-walk framework of Shlesinger-Montroll, we provide a unified molecular interpretation of stretched-exponential relaxation, where the exponent $\beta$ reflects the dimensionality of the underlying conformational dynamics.
\end{abstract}


\maketitle

\section{\label{sec:introduction}Introduction}

Classical polymer physics models, such as the Rouse, Zimm, and tube models (including their many variations), rely on the concept of a uniform and statistically ideal segment to simplify the intricate dynamics of long-chain polymers.\cite{Teraoka-PolymSol-2002,Doi-Edwards-ThePolymDyn-1994,Rubinstein-Colby-PolPhys-2003} However, difficulties arise as soon as numerical values are inserted into these models and the predictions are compared with experimental results.

A basic question lies at the root of these discrepancies: what exactly is a \emph{statistical segment}? What is the minimal size that a segment must have to behave statistically? Common sense suggests that the smallest statistical segment is the Kuhn segment $l_\text{k}$. Yet in practice several different definitions of “statistical segment’’ are used in theoretical and numerical models, and they do not always coincide with the Kuhn segment.\cite{Rubinstein-Colby-PolPhys-2003,Lodge-Heimenz-PolyChem-2007,Larson-McLeish-2003} This ambiguity affects the evaluation of relaxation times, the identification of relevant length scales in flow experiments, and the interpretation of melt dynamics.

It is also well established that short chains are unentangled and display non-Gaussian behavior, whereas chains above a critical molecular mass are entangled and approach Gaussian statistics at the chain level. Gaussianity is commonly assumed to emerge once the chain contains roughly ten statistical segments.\cite{Yamakawa-1971} But this raises an important question: are the Kuhn segments in a chain, whether unentangled or entangled, actually all similar?

All classical theories implicitly assume that they are. The assumption of identical statistical segments leaves no room for any heterogeneity at the Kuhn-segment scale. Yet previous works\cite{Martins-macromol-2013,Kavassalis-Sundararajan-macrom-1993,Sundararajan-Kavassalis-JCSoc-1995} have shown that heterogeneities at this level do exist, although procedures for their identification and characterization have not yet been validated. We address here these two issues. This immediately leads to further questions. What is the role of these heterogeneities in melt relaxation? How relevant are they to early stages of crystallization, to the yield stress observed at the onset of shear, or to the melt memory effect? And are these heterogeneities connected with those observed in polymer melts approaching the glass transition?

The end-to-end distance of a polymer chain in a melt fluctuates in time, giving rise to a distribution of distances. This distribution is Gaussian for sufficiently long, entangled chains, while unentangled chains show clear deviations from Gaussianity. These fluctuations are typically interpreted by decomposing the chain into a sequence of statistical segments, each treated as an entropic spring.\cite{Teraoka-PolymSol-2002,Rubinstein-Colby-PolPhys-2003}

But is it accurate to assign the role of an entropic spring to a single statistical segment? Sokolov concluded that at least five statistical segments are needed for a chain to display Rouse-like behavior in polystyrene melts.\cite{Ding-Sokolov-JApplPolySci-2004} Inoue \textit{et al.}, also for polystyrene, found a correlation between the stress–optical coefficient and the Rouse segment, and estimated the mass of that segment to be about \SI{1000}{g.mol^{-1}}, $\approx 1.4$ times the mass of one Kuhn segment [$M_\text{k} = 700$~\si{g.mol^{-1}}, Table 2.1 in Ref.~\onlinecite{Rubinstein-Colby-PolPhys-2003}].\cite{Ionue-macrom-2002} These two very different results highlight the need to clarify what a statistical segment is, what its minimal size must be, and what the minimal size of an entropic spring is.

For long chains in the melt, the standard model of flow assumes the confinement of a chain within a tube; that is, chain segments explore a constrained region in space due to entanglements.\cite{Doi-Edwards-ThePolymDyn-1994,Dealy-Read-Larson-2018} But to date there exists no molecular-level definition of entanglement. Moreover, recent experimental results showed that even within the entanglement volume, the dynamics of entanglement strands is significantly non-Gaussian,\cite{Kruteva-Zamponi-macom-2021} in contradiction with the assumptions of essentially all models of flow in entangled melts. Such results demonstrate that the physical nature of entanglements is not merely an academic issue, but a central question for constructing realistic models of polymer dynamics.

Another experimental observation adds to this picture. \citeauthor{Wischnewski-2003a} used neutron spin-echo measurements to identify the transition from the unconstrained $t^{1/2}$ regime to the constrained $t^{1/4}$ regime. The mean-squared displacement at which this transition occurs in polyethylene at 509 K is approximately $300$~\AA$^2$, only $\approx 1.5$ times the length of one Kuhn segment (Fig.~2 in Ref.~\onlinecite{Wischnewski-2003a}). This evaluation assumes that the Gaussian approximation is valid at this short length scale.

However, when one inspects the length scale associated with the mean-squared displacements $\langle \Delta r^2(t) \rangle$ from molecular dynamics simulations in the literature,  a significant inconsistency emerges with respect to the experimental value.  Hsu and Kremer, using bead-and-spring models, set the onset of the constrained regime at $\approx d_T$ at $\tau_e$, the tube diameter and entanglement relaxation time, respectively (Fig.~6 of Ref.~\onlinecite{Hsu-Kremer-2016}). Using the data of Fetters \textit{et al}. in Table 25.1 of Ref.~\onlinecite{Fetters-2007}, we find $d_T \approx 36$~\AA\ and a length scale of 6–7 Kuhn segments at the onset of the constrained regime. 

The results of atomistic molecular dynamics simulations of Ramos \textit{et al.}\cite{Ramos-2018} and Harmandaris \textit{et al.}\cite{Harmandaris-2003} fall far from this evaluation. Ramos \textit{et al.} following the recommendations of \citeauthor{Likhtman-2002},\cite{Likhtman-2002} set the onset of the constrained regime at the intersection point between lines with slopes $ 1/2$ and $1/4$. A similar procedure was followed by \citeauthor{Harmandaris-2003}.\cite{Harmandaris-2003} The former evaluated the crossover at a length scale of $\approx 360$~\AA$^2$, around 1.8 Kuhn segments (Fig.~5A of Ref.~\onlinecite{Ramos-2018}), and the evaluation of Harmandaris \textit{et al.} points to a crossover at $\approx 10^{2.3}$~\AA$^2$, approximately one Kuhn segment. Both atomistic results are close to the experimental result of \citeauthor{Wischnewski-2003a}. 

The disagreement between these results and the evaluation of Hsu and Kremer justifies a deeper analysis of the statistics of short-chain segments. They also confirm the conclusions of \citeauthor{Kruteva-Zamponi-macom-2021}\cite{Kruteva-Zamponi-macom-2021} on the non-Gaussianity of the dynamics of the entanglement strands.

Here we address several of the questions raised above. We use atomistic molecular dynamics simulations to examine how chain segments behave at the scale of the Kuhn segment and beyond. The system considered is an entangled polyethylene melt with chains of 3500~\si{g.mol^{-1}}. We divide each chain into segments of different sizes; for each size we evaluate the distribution of distances between the first and last atoms of the segment for all 560 chains in the melt, and fit each distribution to a Gaussian function. With this approach we determine the minimal statistical segment, the minimal entropic spring, the minimal segment whose distribution is well described by a Gaussian function, and the length scale marking the transition to Gaussianity.
We further provide a probabilistic evaluation, supported by a reliability test, to define a limiting boundary separating sequences that are more aligned from others more random. A sequence $g^+tg^-g^+tg^-g^+tg^-$ lies beyond this boundary, and should therefore relax more rapidly than sequences inside it. After establishing these definitions, we analyze the orientational relaxation and translational diffusion of the different segment types. 

In doing so, we reveal that a single Kuhn segment is statistically uncorrelated yet strongly non-Gaussian, that the minimal entropic spring and minimal Gaussian segment each require multiple Kuhn segments, and that Kuhn-scale heterogeneity organizes into distinct dynamical substructures with different relaxation signatures. This provides a direct molecular-level connection between torsional cooperativity, stretched-exponential relaxation, and the emergence of heterogeneous dynamics at the Kuhn length scale.

\section{\label{sec:fundamentals}The Fundamentals}

The dimensions of polymer chains are determined by the mean-squared radius of gyration $\langle R_g^2 \rangle$, measured in scattering experiments. The procedure is described in standard textbooks. \cite{Teraoka-PolymSol-2002,Rubinstein-Colby-PolPhys-2003,Lodge-Heimenz-PolyChem-2007,Dealy-Read-Larson-2018} Under the assumption that the backbone trajectory in space is described by a 3-D random walk, the mean-squared end-to-end distance of the real chain is evaluated  as
\begin{equation}
	\langle R^2 \rangle_{\text{real}} 
	= 6 \langle R_g^2 \rangle 
	= C_{\infty}\, n_{\text{cc}}\, l_{\text{cc}}^2,
	\label{eq:R2-real-chain}
\end{equation}
where $C_\infty$ is the characteristic ratio, $n_{\text{cc}}$ is the number of C--C bonds in the chain (evaluated from the polymer's average molecular weight), and $l_{\text{cc}}$ is the C--C bond length ($\approx1.54$ \AA). Equation (\ref{eq:R2-real-chain}) holds for flexible chains in the Gaussian regime, such as in polymer melts, where excluded-volume effects are screened.\cite{Flory-PPChem-1953,Flory-JCP-1949}

The key parameter in this equation is the \textit{characteristic ratio}. It is defined by IUPAC as the ratio between the experimentally determined $\langle R^2 \rangle$ of polymer chains of known molecular weight under ideal conditions (melts or solutions at the theta temperature) and the product of the number of main-chain bonds and the squared bond length. For vinylic polymers, this product is ($n_{\text{cc}} l_{\text{cc}}^2$). The characteristic ratio encapsulates all intramolecular constraints: bond angles, torsion angles, and the statistical weights of torsional states. The effect of intermolecular interactions is also naturally included in the above evaluation.  Like other polymer physical properties (e.g., density, glass transition temperature and melting temperature), the characteristic ratio increases with molecular weight and approaches a constant value, $C_\infty$, above the critical molecular weight (Figs. 3 and 4 in Ding and Sokolov \cite{Ding-Sokolov-macrom-2004} and Fig. 4 in Fatou and Mandelkern \cite{Fatou-1965}).

Several different lengths have been introduced to describe the random walk representation of real polymer chains. The most common in polymer physics and chemistry are the persistence length $l_\text{p}$, the statistical segment length $b$, the effective segment length $l_{\text{eff}}$, and the Kuhn length $l_{\text{k}}$ (sometimes denoted $b$, $b_\text{k}$, or $l_{\text{k}}$). Relationships among these lengths and with the real chain are provided in textbooks \cite{Gedde-2019,Lodge-Heimenz-PolyChem-2007,Rubinstein-Colby-PolPhys-2003,Dealy-Read-Larson-2018} and discussed in articles. \cite{Larson-McLeish-2003,Ding-Sokolov-JApplPolySci-2004,Ding-Sokolov-macrom-2004}

\subsection{\label{subsec:Kuhn-Segment}The Kuhn Segment}

The Kuhn segment length ($l_\text{k}$) is evaluated by equating the mean-squared end-to-end distance of the real chain with that of an equivalent freely jointed chain containing $N_\text{k}$ segments with length $l_\text{k}$, 
\begin{equation}
	\langle R^2 \rangle_{\text{k}} 
	= N_\text{k}\ l_\text{k}^2 
	= C_{\infty}\, n_{\text{cc}}\, l_{\text{cc}}^2 \; .
	\label{eq:R2-kuhn-chain}
\end{equation}
To fully establish the above relationship, two constraints (or conservation conditions) are imposed: (\textit{i}) the equivalent chain cannot be longer or shorter than the real chain, and (\textit{ii}) both chains must have the same mass. With
\begin{equation}
	R_{\text{max}} = N_\text{k}\ l_\text{k} 
	=  K_{\text{geom}}\, n_{\text{cc}}\, l_{\text{cc}} ,
	\label{eq:R-max}
\end{equation}
we obtain the following well known equations:
\begin{subequations} 
	\begin{align}
		l_\text{k} &= \frac{C_\infty \, l_{\text{cc}}}{K_{\text{geom}}} ,
			\label{eq:l-kuhn} \\
		N_\text{k} &= \frac{n_{\text{cc}} \, K^2_{\text{geom}}}{C_\infty} \, .
		\label{eq:N-kuhn}
	\end{align}
\end{subequations}
The geometrical constraint, $K_{\text{geom}}$, expresses the dependence on the polymer's chemical structure, such as whether the most stable spatial conformation is a planar \textit{zig-zag} (as in polyethylene, PE) or a helical structure.\cite{Dealy-Read-Larson-2018}  For PE $K_{\text{geom}} = \sin(\theta/2) $. 
With $j$ C--C bonds per repeat unit, the mass of a Kuhn segment is
\begin{equation}\nonumber
	M_\text{k} = \frac{C_\infty\, M_\text{ru}}{j K^2_{\text{geom}}}\ ,
	\label{eq:M-Kuhn}
\end{equation}
from which the number of repeat units per Kuhn segment can be evaluated. 

Equations~(\ref{eq:R2-kuhn-chain}) and (\ref{eq:R-max}) ensure that the real chain and its freely jointed equivalent have the same mean-squared end-to-end distance and the same fully stretched contour length. Importantly, these conditions do \emph{not} imply that the equivalent chain must reproduce the local orientation or conformation of the real chain. A freely jointed chain consists of uncorrelated segments with arbitrary spatial orientations. It matches the real chain only in its global statistics. Thus, the Kuhn segment is a statistical construct, not a locally straight or rigid block. This remark is fully consistent with our atomistic results, which show that individual Kuhn segments span a broad range of conformations.

\subsection{\label{subsec:Statistical-b}The Statistical Segment Length}

In some works the Gaussian chain is described using the number of repeat units and an effective step length $b$, sometimes called a ``\textit{monomer-based segment length}'' or ``\textit{statistical segment length}''. \cite{Lodge-Heimenz-PolyChem-2007,Larson-McLeish-2003} This terminology is misleading: the monomer refers to the molecule before polymerization (example: ethylene), whereas the repeating unit of the polymer backbone (–\ce{CH2}–\ce{CH2}– in polyethylene) is the relevant structural motif. Hence, in this work these random walk steps are referred to as chain segments or blocks of $n_{\text{cc}}$ C--C bonds.

The statistical segment length $b$ is defined with an equation similar to Eq.~(\ref{eq:R2-kuhn-chain}), 
\begin{equation}
	\langle R^2 \rangle_{\text{st}} = N_\text{st}\, b^2 = C_{\infty}\, n_{\text{cc}}\, l_{\text{cc}}^2,
	\label{eq:R2-statistical-b}
\end{equation}
with a one-to-one correspondence between the number of steps and the number of repeat units, $N_\text{st} = n_{\text{ru}}$. The number of repeat units is related to the number of C--C bonds ($n_{\text{cc}}$) and the number of C--C bonds per repeat unit ($j$) by $n_{\text{ru}} = n_{\text{cc}}/j$. For polyethylene $j=2$ C--C bonds in the chain backbone. 
The number of statistical steps can also be expressed in terms of the chain mass ($M_\text{ch}$) and the repeat unit mass ($M_\text{ru}$) as $N_\text{st} = M_\text{ch}/M_\text{ru}$. This makes $b$ a \textit{repeat-unit-based length} with
\begin{equation}
	b = \sqrt{j\,C_\infty}\, l_{\text{cc}}\, 
	\label{eq:b-definition}
\end{equation}
for $j$ C--C bonds per repeat unit. The relationship between $b$ and the Kuhn segment is evaluated by requiring that both representations (the repeat-unit-based and Kuhn) should yield the same mean-squared end-to-end distance:
\begin{equation}\nonumber
	N_\text{st}\, b^2 = N_\text{k}\, l_\text{k}^2 \,.
\end{equation}
This last equation provides a way to express $\langle R^2 \rangle$ in terms of repeat units, allowing the conversion to Kuhn segments. \cite{Lodge-Heimenz-PolyChem-2007,Larson-McLeish-2003} For polyethylene chains,
\begin{subequations}
	\begin{align}
	N_{\text{k}} = \frac{j\, n_{\text{ru}}\, \sin^2(\theta/2)}{C_\infty}, \\
	l_\text{k} = \frac{\sqrt{C_\infty/j}}{\sin(\theta/2)}\, b.
	\label{eq:lk-from-b-Nk-from-n_ru}
	\end{align}
\end{subequations}
The geometrical factor $K_{\text{geom}} = \sin(\theta/2) $ constrains the most extended spatial arrangement of the chain to an all-\textit{trans} conformation.

\subsection{\label{subsec:Static-Dynamic-Segments} Static and Dynamic Segments Definitions}

The above statistical step length $b$, as well as $l_\text{k}$, are considered static segments. To define entropic springs, \citeauthor{Ding-Sokolov-JApplPolySci-2004}  \cite{Ding-Sokolov-macrom-2004} proposed rescaling the number of statistical steps based on an entropic spring segment, $N_{\text{st}} = M_\text{ch}/M_\text{es}$. In this picture, $b$ plays the role of a dynamic bead size (the smallest subchain that still exhibits the Rouse-like dynamics), typically much larger than one Kuhn segment (dynamic bead length: 45 \AA, $l_\text{k} = 14$  \AA, for polyethylene (PE)). 

\citeauthor{Agapov-Sokolov-macrom-2010} considered also that the  static bead size is inconsistent with the Kuhn segment length.\cite{Agapov-Sokolov-macrom-2010} They argued that ``the deficiency of the traditional definition of the Kuhn segment ... is based on the assumption of the chain being completely extended (all \textit{trans} conformations in the case of PE) inside a single bead''. 

We note that the term  ``bead'' belongs to their dynamic interpretation of segmental motion; in the present work all quantities are evaluated strictly at the atomistic level, and no bead or coarse-grained representation is introduced. The direct atomistic evaluation of intrachain distances associated with the Kuhn segment helps clarify precisely the issue raised in the Agapov and Sokolov's argument.

\subsection{Gaussian Fits and Fit Quality}

In this work we analyze the end-to-end distances' distribution in blocks of C--C bonds with different sizes, starting with half of the Kuhn segment up to large fractions of the full chain length, restricting the analysis to PE chains with $M>M_\text{c}$. The distribution of end-to-end distances for each block was normalized and fitted to a Gaussian distribution function
\begin{equation}
	P(R) = 4 \pi R^2 \left( \frac{3}{2 \pi N_{\text{step}}\, l_{\text{step}}^2} \right)^{3/2}
	\exp \left(- \frac{3 R^2}{2 N_{\text{step}}\, l_{\text{step}}^2}\right),
	\label{eq:radial-gauss-dist-function}
\end{equation}
where $N_{\text{step}}$ is the imposed number of Gaussian steps used in the fragmentation of the chain and $l_{\text{step}}$ the fitting variable. The product $N_{\text{step}} l_{\text{step}}$ must satisfy the constraint of Eq.~(\ref{eq:R-max}). By analyzing the quality of the fits, applying additional gaussianity tests, and examining quantile--quantile (Q-Q) plots, we determine the minimal number of C--C bonds required to define a statistical segment, and the minimal number of such segments required to define an entropic spring. In addition, we also analyze the  orientational relaxation and translational dynamics associated with different types of Kuhn segments.

\section{Materials and Methods}
\label{sec:methods}
\subsection{Simulation Details}
\label{sec:simulation-details}

Molecular dynamics simulations were performed following the protocol described in a 
previous work.\cite{Martins-macromol-2013} The polyethylene (PE) model is that of 
\citeauthor{Paul-1995a},\cite{Paul-1995a,Paul-1997a} with chains assembled into a cubic box of $\approx 160^3\; \text{\AA}^3$ at $\rho = 0.688\;\si{g.cm^{-3}}$. The simulated melt contained 140,000 united atoms, corresponding to 560 chains of 250 atoms each, and was equilibrated in the isothermal–isobaric ensemble at $T=600$ K and $p=1$ atm. 

The chain mass is above the critical mass of PE,\cite{Dealy-Read-Larson-2018} 
with $2M_\text{e} \leq M_\text{c} \leq 3M_\text{e}$ and $M_\text{e} \approx 1200 \,\si{g.mol^{-1}} $. All simulations were performed using GROMACS 4.5.3.\cite{Hess-2008a} Further methodological details are provided in the Supplementary Material (Section S1 and Table S1).

\subsection{Grid Search, Fitting, Constraints, and Acceptance Rules}
\label{sec:grid-search}

A grid search is performed over block sizes $n_{\text{cc}}$  (number of C--C bonds, e.g., 6, 11, 17, \ldots, 249  bonds) and possible divisions $N_{\text{steps}}$ (e.g., 1, 2, 5, \ldots, 50 steps).  For each ($n_\text{cc}, N_\text{steps}$) pair, the normalized distribution of the end-to-end distances within each block  was fitted to the Gaussian function in Eq.~(\ref{eq:radial-gauss-dist-function}) with $l_\text{step}$ as fitting variable. This fit was subjected to two physical constraints: (\textit{i}) a constraint on $N_{\text{steps}}$, and (\textit{ii}) a constraint on the maximum fitted step length.

Before proceeding, it is imposed a clarification on the notation. The symbol $b$ is reserved for the conventional ``statistical segment length'', defined based on the repeat unit, Eq. ~(\ref{eq:b-definition}). To avoid confusion, the length parameter obtained from the Gaussian fitting procedure in Eq.~(\ref{eq:radial-gauss-dist-function}) is denoted $l_\text{step}$.

The minimum allowed value for $N_{\text{steps}}$ is one (the entire $n_\text{cc}$ block, regardless its number of C--C bonds, is treated as a single segment).  The $N_{\text{steps}}$ maximum value is determined by requiring that the length of the smallest segment in the block exceeds the chain’s persistence length (a statistical segment cannot have orientational correlations). In addition, the fitted step length $l_{\text{step}}$ is limited to the maximum physically allowed length for a block of $n_{\text{cc}}$ bonds divided into $N_{\text{steps}}$ steps:
\begin{equation}
	l_{\text{step}} \leq l_{\max} = \frac{n_{\text{cc}}}{N_{\text{steps}}}\, l_{\text{cc}} \sin\!\left(\frac{\theta}{2}\right).
\end{equation}
Depending on the acceptance rules described below, the search may terminate before testing the persistence-length lower bound.

\subsubsection{Data Characterization}

For each block size $s \equiv n_{\text{cc}}$, the normalized histogram of end-to-end distances is characterized by its mean $\langle R \rangle$, variance $\langle R^2 \rangle$, and higher moments that quantify deviations from Gaussianity. \cite{Stanford-QQ-plots-1994,Sivia-Skilling-2006} 

The skewness ($\gamma_1$), or the third standardized moment, measures the asymmetry of the distribution:
\begin{equation}
		\gamma_1 = \frac{\langle (R-\langle R\rangle)^{3}\rangle}{\langle (R-\langle R\rangle)^{2}\rangle^{3/2}}.
		\label{eq:skewness}
\end{equation}
	Positive values ($\gamma_1 > 0$) indicate that the right tail is longer that the left one. The opposite holds for ($\gamma_1 < 0$). 
	 
The excess kurtosis ($\kappa_{\mathrm{ex}}$), the fourth standardized moment, quantifies the extension of the tails relative to a Gaussian:
\begin{equation}
		\kappa_{\mathrm{ex}} = \frac{\langle (R-\langle R\rangle)^{4}\rangle}{\langle (R-\langle R\rangle)^{2}\rangle^{2}} - 3.
		\label{eq:kurtosis}
\end{equation}
Positive values correspond to larger tails, whereas negative values correspond to smaller tails.
	
Deviations from ideal chain behavior are measured by the non-Gaussianity parameter ($\alpha_2$):
\begin{equation}
		\alpha_2 = \frac{3}{5}\,\frac{\langle R^4\rangle}{\langle R^2\rangle^{2}} - 1.
		\label{eq:alpha-2}
\end{equation}
A value of zero corresponds to an ideal Gaussian chain. To account for statistical noise, systems with $|\alpha_2| \lesssim 0.05$ are considered also Gaussian. The following boundaries are practical choices: $0.05 \leq |\alpha_2| < 0.1$ defines a transition zone, and  $0.1 \leq |\alpha_2| < 0.3$ indicates mild non-Gaussianity, with deviations from 10\% and 30\%, respectively, from  $\langle R^4\rangle/\langle R^2\rangle^{2} = 5/3$. Larger $\alpha_2$ values  correspond to high non-Gaussianity. Chains are more compact for negative values and more extended for positive values of $\alpha_2$
	
Finally, the chain flexibility is characterized by the apparent mean-squared segment length per bond
\begin{equation}
	C_{\text{app}}(s) = \frac{\langle R^{2}(s)\rangle}{s},
	\label{eq:c_app-R^2(s)/s}
\end{equation}
with units of \AA$^2$/bond. Its plateau value at large $s$ is related to $l_k^2$
\begin{equation}
	l_k^2 = C_{\text{app}}^{\text{plateau}} \times n_{\text{cc, per Kuhn}},
	\label{eq:c-app-lk}
\end{equation}
where $n_{\text{cc, per Kuhn}}$ is the number of C–C bonds per Kuhn segment. At small $s$, $C_{\text{app}}(s)$ reflects interactions within this segment. As $s$ increases, local interactions fade away, the statistics approach a random walk, and $C_{\text{app}}(s)$ converges to a plateau.\cite{Jayaraman-macrom-2019,Everears-Grest-R2/r-JCPhys-2003}

\subsubsection{Fit Quality Characterization}

For each $(n_{\text{cc}}, N_{\text{steps}})$ pair, the Gaussian fit is evaluated using metrics from both histogram and Q–Q plot analyses.\cite{Riso-Statistical-wR-2008} For the data histogram fit to Eq.~(\ref{eq:radial-gauss-dist-function}), we compute the root-mean-square error ($\mathrm{RMSE}_{\text{hist}}$) and coefficient of determination ($R^2_{\text{hist}}$); for the Q–Q plot, we compute the corresponding quantities ($\mathrm{RMSE}_{\text{QQ}}$) and ($R^2_{\text{QQ}}$).

\subsubsection{Acceptance Criteria}

A fit is considered acceptable only if all of the following criteria are satisfied.
\paragraph*{Physical Constraint.}
The physical constraint requires that the fitted step length does not exceed the maximum physically allowed value $l_{\text{step}} \leq l_{\max}$.

\paragraph*{Statistical Goodness.}
The statistical goodness-of-fit is accessed via the  Q–Q plot, requiring a sufficiently high correlation coefficient to indicate Gaussianity, $ R^2_{\text{QQ}} \geq 0.98$.

\paragraph*{Histogram Fit.}
The histogram fit should match the observed distribution within the error threshold $ \mathrm{RMSE}_{\text{hist}} \leq 0.01$.  
	
These above procedures establish, for each block size $n_{\text{cc}}$, the range of divisions in $N_{\text{steps}}$ that yield a statistically valid Gaussian representation under the imposed physical constraints.

\subsection{The Local Conformational Domain and ACS/RCS identification}

For each Kuhn segment (a block of $n_{\text{cc}}$ backbone bonds; $n_{\text{cc}} \approx 11$ for PE), a local axis is defined as the line connecting the first and last carbon atoms of the block. The perpendicular distance $d$ of each backbone atom to this axis is computed. The maximum value of $d$ across all segments defines the Kuhn segment diameter, $d_k$, confining all chain atoms within a limiting boundary associated to a critical  $r_\text{k}=d_\text{k}/2$ around the local axis of the Kuhn segment.

To classify segments based on their conformation, a \textit{local conformational domain} was defined. Aligned chain segments (ACS) were identified by introducing a critical radius $d_{\text{crit}}$,
\begin{equation}
	d_{\text{crit}} = (1-e^{-1})\,r_k.
	\label{eq:critical-distance-dk}
\end{equation}
Atoms within $d_{\text{crit}}$ are assigned to ACS. Segments outside this criterion and located between two ACS segments are classified as random conformational sequences (RCS). Segments extending from an ACS to the chain end are designated as chain ends (CE). Sections \ref{sec:results-distance-distribution} and \ref{sec:reliability} present a probabilistic justification for the $d_{\text{crit}}$ criterion, including a reliability analysis.

\section{Results and Discussion}
\label{sec:results}

\subsection{Chain Dimensions and Characteristic Lengths}
\label{sec:results-chain-dimensions}

Molecular dynamics simulations offer an important advantage over scattering experiments for evaluating chain dimensions, since they allow the direct measurement of the mean-squared end-to-end distance and the verification of the relation in Eq.~(\ref{eq:R2-real-chain}). For chains of length C250 and longer the ratio obtained is $\langle R^2 \rangle/\langle R_g^2 \rangle \approx 6$, in agreement with the expected value for  Gaussian chains (Fig. 1.C in Ref. \onlinecite{Martins-macromol-2013}). 

\begin{figure}[h!]
	\centering
	\includegraphics[width=0.45\textwidth]{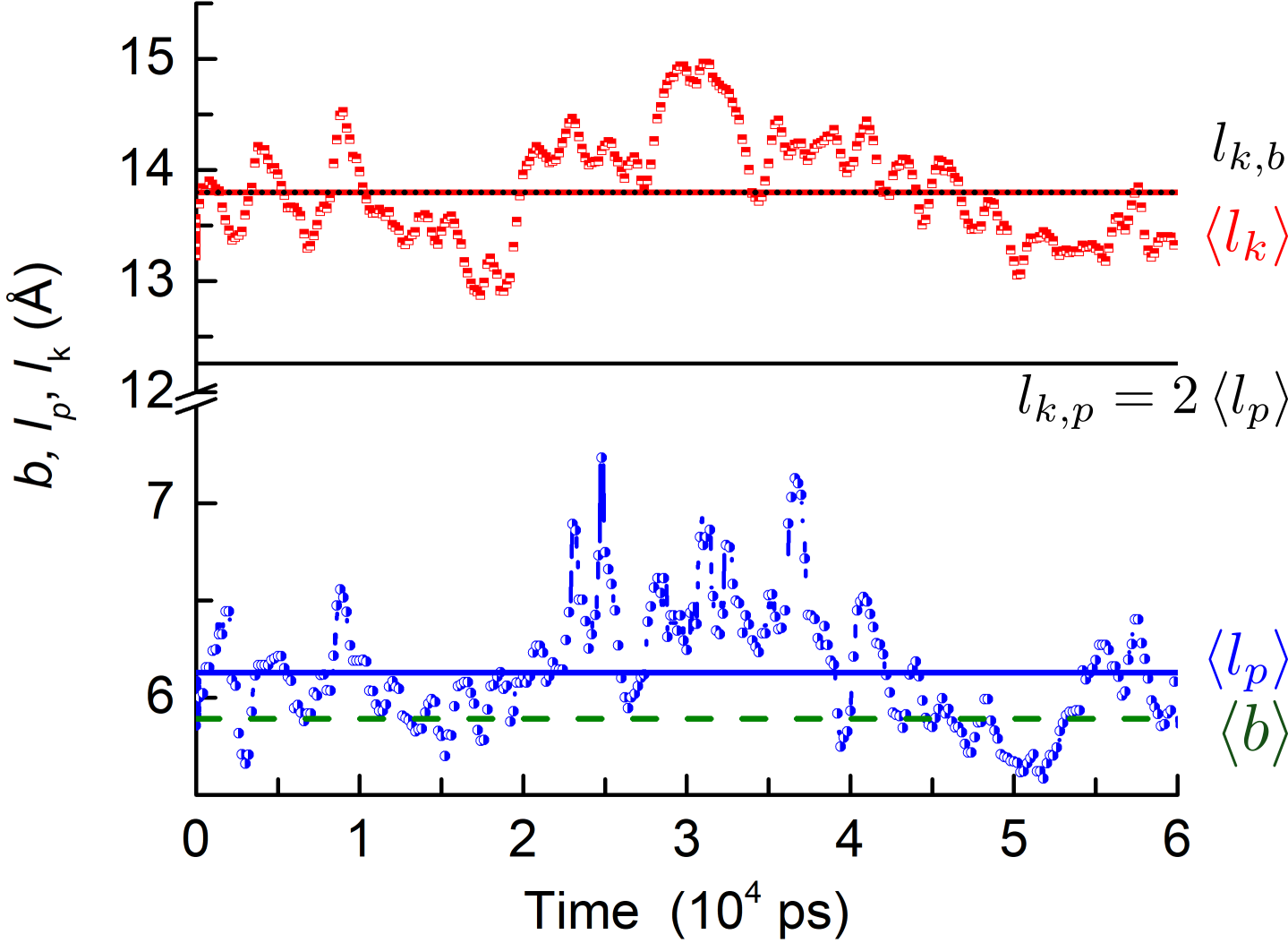}
	\caption{Average segment lengths obtained for the C250 system (560 chains) at $T=600$ K taken over $6.0\times 10^4$ ps. The characteristic ratio is $C_\infty = 7.32$, giving $l_\text{p} = 6.13$ \AA, $b = 5.89$ \AA~ [Eq.~(\ref{eq:b-definition})], and $l_k = 13.80$ \AA~[Eq.~(\ref{eq:l-kuhn})] . Kuhn segment molecular weight: $M_\text{k} = 153.901 \pm 5.038$ \si{g\, mol^{-1}}. The consistency between $l_k$ values obtained by different procedures demonstrates the reliability of the molecular dynamics results.}
	\label{fig:Figure-all-lengths}
\end{figure}

These measurements are used to determine the characteristic ratio $C_\infty$, enabling the calculation of the statistical segment length $b$, and the Kuhn length $l_k$ from the definitions in Eqs.~(\ref{eq:R2-statistical-b} ) and (\ref{eq:R2-kuhn-chain}), respectively. The persistence length is evaluated directly from the bond vector correlations.  Figure \ref{fig:Figure-all-lengths} summarizes the results obtained for the C250 system at $T=600$ K. The agreement between $l_k$ obtained from the statistical segment length [Eq.~(\ref{eq:lk-from-b-Nk-from-n_ru})] and that directly calculated from $\langle R^2 \rangle$ [Eq.~(\ref{eq:l-kuhn})] confirms the internal consistency of the analysis. However, for flexible polymer chains, the ratio $l_\text{k}/l_\text{p}$ is not two. A possible explanation is that bond orientations in flexible chains decorrelate more rapidly than in semiflexible chains. 

\subsection{Definition of Statistical Segment}
\label{sec:results-statistical-segment}

A question naturally emerges from the data in Fig. \ref{fig:Figure-all-lengths}: among the lengths displayed, which can be properly classified as \textit{statistical}?

A \textit{statistical segment} is  defined here as the smallest chain segment for which a model built by concatenating a finite number of such segments, relevant to the system under study, satisfies the following conditions:
\renewcommand{\labelenumi}{(\roman{enumi})}
\begin{enumerate}
	\item The end-to-end distance distribution of a chain formed by the segments exhibits Gaussian statistics, following Eq.~(\ref{eq:radial-gauss-dist-function}).
	\item The central limit theorem applies: correlations between internal degrees of freedom (bond angles, torsional states) are sufficiently averaged out, so that the segments behave as independent random vectors.
	\item Successive segments are, to a good approximation, uncorrelated in orientation (beyond persistence length effects).
\end{enumerate}
\renewcommand{\labelenumi}{\arabic{enumi}.} 
This practical definition aims to move beyond the Gaussian limitation of an infinite number of segments, thus ensuring the model validity under the constraints described in the Methods Section.  

Condition (iii) excludes the persistence length $l_p$, which measures orientational correlations rather than Gaussian behavior. The data in Fig. \ref{fig:Figure-all-lengths} also excludes $b$ from the statistical classification.

This is an important result, inasmuch as it conflicts with recommendations issued by a prominent group of authors (\citeauthor{Larson-McLeish-2003}\cite{Larson-McLeish-2003}) to correct errors and inconsistencies in comparisons between experimental results and quantitative predictions of the tube model. According to these authors, the main source of error is the incorrect use of one of three different definitions of the tube model parameters (Graessley, Ferry and Milner-MacLeish, Table I in Ref. \onlinecite{Larson-McLeish-2003}). In this recommendation, the equations for the plateau modulus and relaxation times are all written in terms of the ``\textit{monomer-based segment length}'' or ``\textit{statistical segment length}''. The basic statistical step-length they considered, as it appears in the \citeauthor{Doi-Edwards-ThePolymDyn-1994} book,\cite{Doi-Edwards-ThePolymDyn-1994} is not statistical at all.

\subsection{Failure of Short Blocks to Meet Statistical Criteria}
\label{sec:results-failure-short-blocks}

To illustrate, to obtain an average step length of $b \approx 6$ \AA~(Table 6.1 in Lodge and Heimenz, Ref.~\onlinecite{Lodge-Heimenz-PolyChem-2007}) requires blocks of about 5--6 C--C bonds. At this scale the number of possible conformations is limited (e.g.\ $3^4 = 81$ for six bonds), insufficient for Gaussian statistics. These blocks are shorter than the orientational correlation length ($l_p\approx 6.13\,\text{\AA}$), so successive 6-bond blocks retain appreciable orientational correlation. Thus, independence between neighboring steps is not achieved. Also, as shown in Fig.~\ref{fig:prob-hist-Q-Q-ncc-6}(a) and Table S3 in the Supplementary Material, negative $R^2$ values were obtained both for the fit of the Gaussian function to the histogram data and for the Q-Q plot [Fig.~\ref{fig:prob-hist-Q-Q-ncc-6}(b)]. In particular the negative RMSE$_{\text{hist}}$ indicates a fit worse than the simple average of the data. Further, the distribution is skewed (skewness $\approx -0.3$), have an excess kurtosis ($\approx-0.33$), and the non-Gaussianity parameter is $\alpha_2 = -0.377$.

\begin{figure*}
	\centering
	\includegraphics[width=0.95\textwidth]{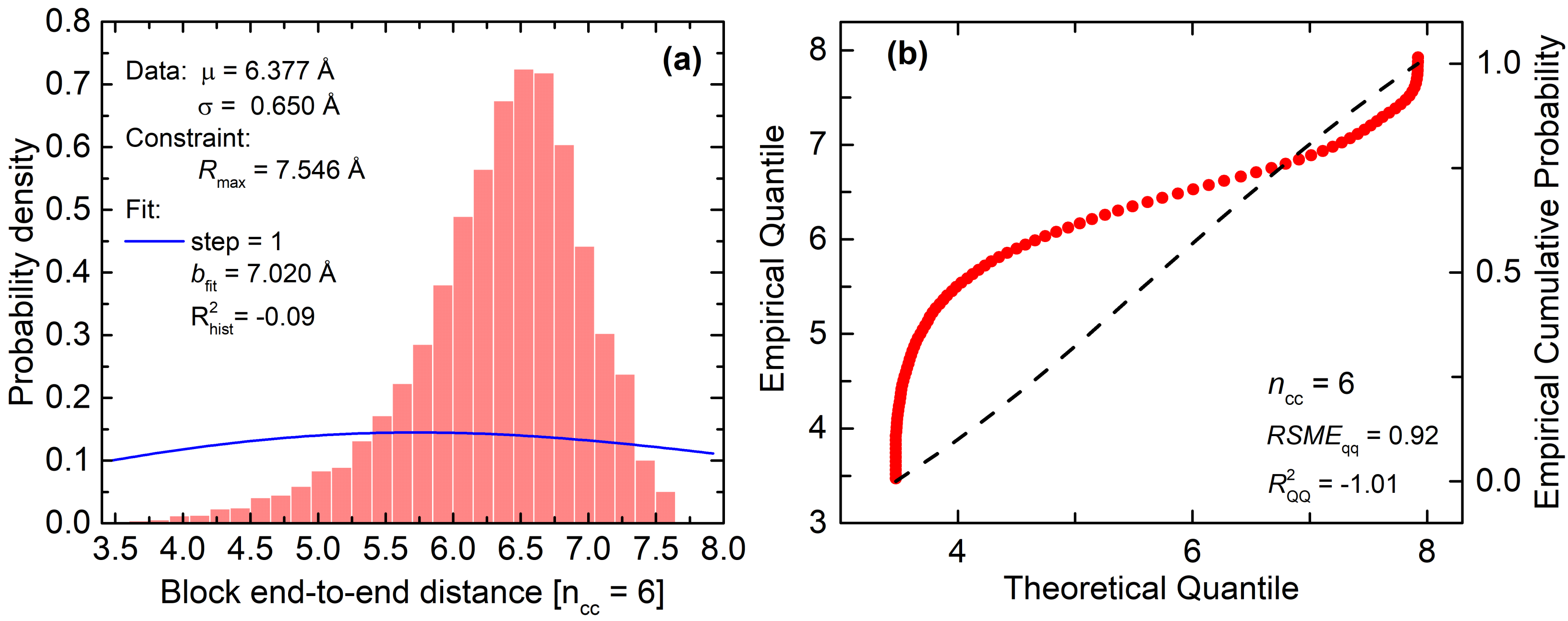}
	\caption{End-to-end internal distance distribution for a segment with $n_\text{cc} = 6$} C--C bonds (a) and Q-Q plot (b). In (b) the data points at the lower end are more spread out than expected (above the line), which is characteristic of a  left-skewed distribution (there are more small values than in a normal distribution).   
	\label{fig:prob-hist-Q-Q-ncc-6}
\end{figure*}

Although the geometric constraint is satisfied, a 6-bond block fails criteria (i)--(iii): its distance distribution is non-Gaussian, internal correlations are not averaged out, and neighboring blocks are not orientation-independent. Therefore, $n_{\text{cc}}=6$ \textit{does not} qualify as a statistical segment. Larger blocks (tested below) are required before Gaussian statistics and effective-step independence emerge.

\begin{figure*}
	\centering
	\includegraphics[width=0.95\textwidth]{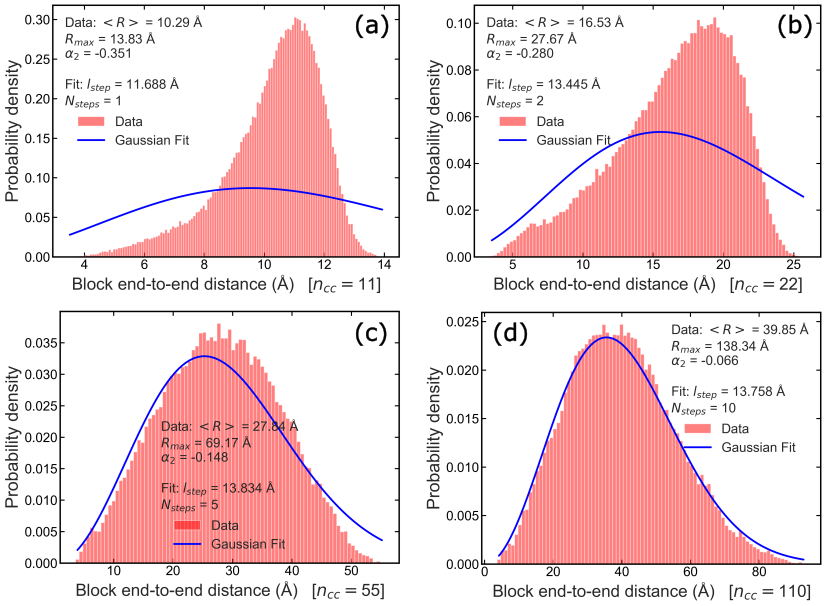}
	\caption{End-to-End distance distribution and Gaussian fit for different $n_\text{cc}$  blocks:  (a) 11 C--C bonds, 1 step; (b) 22 bonds, 2 steps; (c) 55 bonds, 5 steps and (d) 110 bonds and 10 steps. The number of bonds of the step size equals those of one Kuhn segment. } 
	\label{fig:probability-blocks-11-to-110}
\end{figure*}

\subsection{Gaussianity of Blocks Containing Kuhn Segments}
\label{sec:results-Gaussianity-of-blocks}

For each $n_{cc}$, the maximum number of steps $N_{\text{steps}}^{\max}$ or the shortest segment that satisfies the acceptance rules was evaluated. Figure \ref{fig:probability-blocks-11-to-110} shows the results for blocks corresponding to one Kuhn segment (11 C--C bonds) up to 10 Kuhn segments (110 C--C bonds). These results are complemented by the Q--Q plots and data in the Supplementary Material (Section S2, Figure S1 and Table S3). The Table S3 summarizes the results obtained with the grid search for the minimal statistical segment, including fit quality and acceptance criteria.

The block corresponding to one Kuhn segment (containing 11 C–C bonds) was fitted with Eq.~(\ref{eq:radial-gauss-dist-function}). This yielded a positive value,
$R^2_\text{hist} = 0.132$ (see Table S3 in the Supplementary Material) and a step length consistent with the physical constraint acceptance criteria. The fitted step length ($l_\text{step} = 11.688$ \AA) is higher than the persistence length, confirming that successive segments are uncorrelated. Hence, conditions (ii) and (iii) are satisfied. Therefore, a single Kuhn segment satisfies the criteria for being the minimal statistical segment: it is an independent unit of motion.

\begin{figure}[h!]
	\centering
	\includegraphics[width=0.48\textwidth]{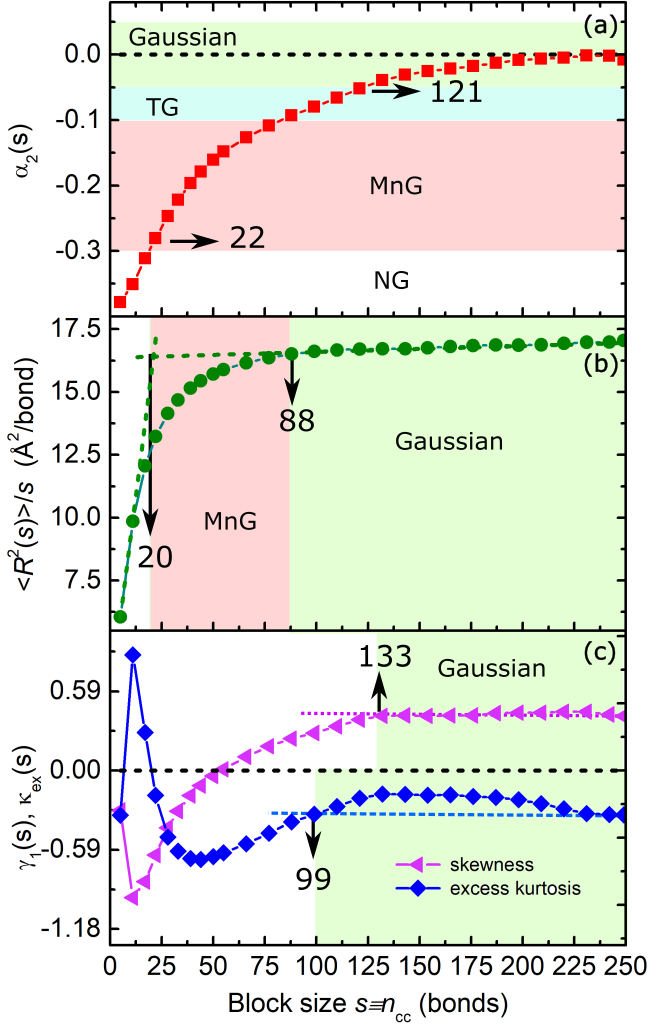}
	\caption{Statistical variables and high-order moments used to characterize the segment-size crossover to Gaussian statistics,  plotted as a function of the block size $s \equiv n_{\text{cc}}$. The first four data points are those shown in Table S3 in the Supplementary Material (6, 11, 17 and 22). (a) The non-Gaussian parameter, $\alpha_2(s)$: The regions indicated, Gaussian, transition to Gaussian (TG), mild non-Gaussian (MnG), and non-Gaussian were defined based on the criteria described in the Methods Section. (b) The apparent segment mean-squared displacement per bond, $\langle R^2(s)/s\rangle$:This quantity probes the characteristic segment stiffness. The interception of two tangent lines, to the plateau and to the initial data, is used to define the onset of mild non-Gaussian behavior $n_\text{cc}\approx20$. (c) Skewness and excess kurtosis: Tangent lines to the data at large block sizes, and large number of steps, are used to define the onset of Gaussinity.}
	\label{fig:kurt-skew-R2-over-s-gama-2}
\end{figure}

However, this single statistical segment is itself non-Gaussian. This is demonstrated by the negative $R^2_{\text{QQ}}$ values (Fig. S1 and Table S3  in the Supplementary Material), and by the additional results in Fig.~\ref{fig:kurt-skew-R2-over-s-gama-2}. The non-Gaussian parameter  $\alpha_2 = -0.351$, the apparent mean-squared segment length per bond $R^2(s)/s$, skewness and kurtosis, all indicate non-Gaussianity behavior. Thus, based on these assessments, a model based on a single Kuhn segment as an entropic spring would be inaccurate for describing short-chain phenomena. 

\subsection{Crossover to Gaussian Statistics}
\label{sec:results-cross-over-to-gaussian}

As expected, block sizes above the Kuhn segment are all statistical. The question regarding Gaussianity depends on the criteria used. If the classification is based only on the acceptance criteria illustrated in Table S3 in the Supplementary Material, the shortest Gaussian segment contains three Kuhn segments (33 $n_\text{cc}$ bonds). If it is based on the data characterization, the minimum value depends on the variable used and on the desired level of Gaussianity.

With relaxed criteria, mild non-Gaussian segments could be accepted. Under these criteria, based on the values of $\alpha_2$ and $C_{\text{app}}(s)$, the apparent mean-squared segment length per bond, the shortest segment that can act as an entropic spring, at least approximately, should contain a minimum of two Kuhn segments (22 $n_\text{cc}$ bonds). Thus, the minimal entropic spring size is constrained by the statistical requirement for Gaussianity.

For stricter criteria, such as $|\alpha_2|\lesssim 0.05$ and the plateau onset of $C_{\text{app}}(s)$, the minimum value is twelve or eight Kuhn segments, respectively. For the results in Fig. \ref{fig:kurt-skew-R2-over-s-gama-2}(c), Gaussian statistics emerge between 130 and 250 bonds, where both skewness and kurtosis approach and stabilize near their Gaussian values. The most dramatic non-Gaussian effects occur below 50 bonds. This value agrees with the \emph{dynamic bead} size reported by Sokolov for PE ($\approx55$ C--C bonds or $\approx45$ \AA).\cite{Agapov-Sokolov-macrom-2010}
As shown in Table S3, the best fits to a Gaussian function are obtained for $n_\text{cc} \ge 55$ where $ \mathrm{RMSE}_{\text{hist}} \le 0.003$. However, based on the higher-order central moments and other non-Gaussian metrics, such as $\alpha_2$ and the apparent mean-squared segment length per bond, described in Fig.~\ref{fig:kurt-skew-R2-over-s-gama-2}, these short segments exhibit mild non-Gaussian behavior, and therefore cannot be classified as perfect entropic springs.

In summary, Fig. \ref{fig:probability-blocks-11-to-110} shows the  distance distribution for the smallest \emph{statistical segment} [Fig.~\ref{fig:probability-blocks-11-to-110}(a)], the smallest entropic spring, although non-Gaussian [Fig.~\ref{fig:probability-blocks-11-to-110}(b)], the shortest segment whose distribution is well-approximated by a Gaussian function [Fig.~\ref{fig:probability-blocks-11-to-110}(c)], and the smallest Gaussian segment containing 10 Kuhn segments [Fig.~\ref{fig:probability-blocks-11-to-110}(d)], a result consistent with classical expectations but now quantified through direct molecular-level criteria.

\subsection{Distance Distribution and Identification of ACS, RCS, and CE}
\label{sec:results-distance-distribution}

\begin{figure*}[t]
	\centering
	\includegraphics[width=0.7\textwidth]{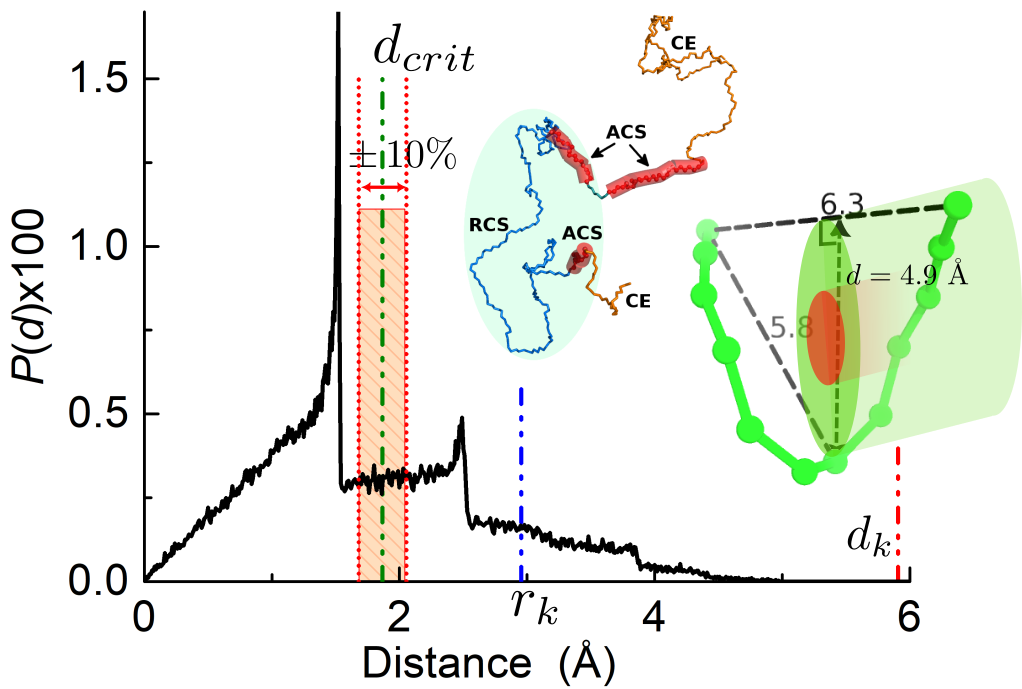}
	\caption{Distribution of perpendicular distances $P(d)$ of backbone atoms to the local Kuhn segment axis in a polyethylene chain. The Kuhn confinement radius $r_{\text{k}}$
	and the critical distance $d_{\text{crit}} = (1-e^{-1})\,r_{\text{k}}$  (with $\pm10\%$ variation band) used to define the inner confinement domain are indicated. Insets: (Bottom right) Schematic of the distance measurement for a helical conformation formed by one Kuhn segment. The green confinement surface confines all atoms of the chain. (Center) Chain snapshot illustrating Aligned Chain Segments (ACS, inside the inner-domain), Random Conformational Sequences (RCS) and Chain Ends (CE) (outside), and a single non-Gaussian Kuhn segment. The normalization factor is the number of atoms in the segments. }
	\label{fig:dk-rk-bands-10-per-cent-inkscape}
\end{figure*}

The identification of dynamically distinct Kuhn segments is based on a geometric analysis of local chain collinearity. Starting at one end of the chain, a local axis is defined as the vector connecting the first and last atom of one Kuhn segment. The perpendicular distance $d$ from this axis to every backbone atom within the window is computed. The window slides along the chain in steps of half a Kuhn segment, and the perpendicular distances are evaluated at each step. The resulting normalized distribution, $P(d)$, is displayed in Fig. \ref{fig:dk-rk-bands-10-per-cent-inkscape}. 

The inset in the lower right of this figure illustrates the procedure. It represents one Kuhn segment, containing 11 C--C bonds and 9 conformations. The conformation shown is a helix, formed by a sequence of alternating \textit{gauche-plus} and \textit{gauche-minus} bonds. The vertical distance from the axis to an atom at the middle of the segment is 4.9 \AA\ in this case. This distance is evaluated for all atoms along the chain, yielding the $P(d)$ distribution and the ultimate value for $d_\text{k}= 5.9$ \AA. Hence, a local conformational domain of radius $r_\text{k}$ (green confinement surface in the inset) would confine all atoms of the chain.

The $P(d)$ distribution in Fig. \ref{fig:dk-rk-bands-10-per-cent-inkscape} highlights that preferred confinement is much tighter: a pronounced peak at 1.52 \AA\ (a vertical distance to the segment axis, not the C--C bond length), a plateau between 1.56 and 2.3 \AA, and a peak at 2.5 \AA. Beyond this point, the decay of $P(d)$ is approximately exponential, with very low probability configurations approaching the ultimate Kuhn diameter $d_k$. These rare events are discussed in the Supplementary Material (Section S3). For consistency and reproducibility, we adopt the axis connecting the atoms in extreme positions of the segment as reference, which ensures that $r_k$ describes the limiting radius.

For a perfectly aligned rigid rod $P(d)$ would be described by a delta function at $d=0$, whereas for a real chain the spread is structured rather than uniform.  Figure~\ref{fig:dk-rk-bands-10-per-cent-inkscape} shows clearly that there is a high--probability region, concentrated near the axis, forming an ``aligned core'', that should also contain  the more extended configurations of the Kuhn segments presented in Fig. \ref{fig:probability-blocks-11-to-110}(a). Details on the specific conformations of these segments are presented below in this work. The plateau and secondary peak indicate a ``semi--ordered shell'', and the exponential tail defines the ``disordered periphery''.

\subsection{Reliability of the Inner-Domain Criterion}
\label{sec:reliability}

A natural step forward is defining a inner confinement domain to identify Kuhn segments in the aligned core. To capture Kuhn segments that contribute significantly to local orientational order and slow dynamics, a definition based solely on the primary peak of $P(d)$ is insufficient, as it excludes semi-ordered states that are still aligned. The most probable atomic positions, with the ability to induce orientational order, can be defined by a characteristic length scale of the spatial spread, which can be used to define the inner confinement domain radius (red cylinder in Fig. \ref{fig:dk-rk-bands-10-per-cent-inkscape}). This is found by evaluating the the cumulative probability $P(d)$ from $d=0$ up to $d_{\text{crit}}$, where it reaches $(1-e^{-1})=63.2\%$.

This criterion sets the critical distance as defined by Eq.~(\ref{eq:critical-distance-dk}). This choice provides a reproducible approach for separating oriented conformational sequences (Aligned Chain Segments) from more random ones (Random Conformational Sequences), without relying on arbitrary thresholds. 

The inset at the center of Fig. \ref{fig:dk-rk-bands-10-per-cent-inkscape} shows the result of this definition after identifying the atoms located inside the inner-domain and applying an additional restraint. To identify different types of Kuhn segments a constraint was set: each segment, at least, should contain the number of atoms found in one Kuhn segment. Too short aligned sequences were discarded. Segments between two ACS were classified as RCS, and should behave as entropic springs. A single Kuhn segment between two ACS, with atoms lying outside the inner confinement domain, is not an entropic spring, and remains unclassified. Segments between ACS and chain ends are, naturally, chain ends (CE). 

Although developed for PE, the procedure is fully general.  
Applying the same analysis to atactic polystyrene (Supporting Material, Fig.~S2) yields an analogous inner-domain structure, demonstrating that the criterion is not limited by side-group size or local sterics.

\begin{figure}[h!]
	\centering
	\includegraphics[width=0.45\textwidth]{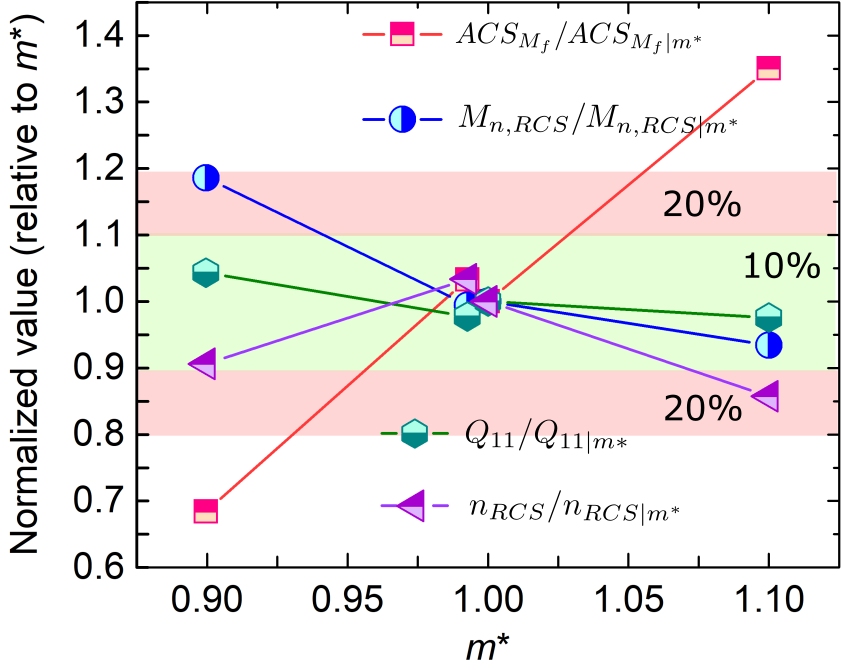}
	\caption{Sensitivity of structural parameters to the inner-domain cutoff 
		$m$. Values are normalized to their baseline at $m^\star = d_\text{crit}/r_{\text{k}} (1-e^{-1})$. The markers represent the baseline ($m^*)$, a cutoff derived from the packing length $m_p = d_\text{crit,p}/r_{\text{k,p}}=(1-e^{-1})$, and $\pm 10\%$ variations of $m^*$. The horizontal bands indicate $\pm 10\%$ and $\pm 20\%$ deviations.}
	\label{fig:normalized-vs-m-star}
\end{figure}

The shaded bands in Fig. \ref{fig:dk-rk-bands-10-per-cent-inkscape} illustrate the inner-domain criterion, $d_{\mathrm{crit}} = (1-e^{-1})\,r_k$, together with the $\pm 10\%$ variation band in the ``semi--ordered shell''. The reliability of this criterion was verified by varying $d_\text{crit} = m\, r_{\text{k}}$ by $\pm 10\%$ around the baseline value $m^\star =  (1-e^{-1})$. We examined the sensitivity to these variations of key structural parameters: the ACS mass fraction, the number-average mass of RCS segments ($M_{\text{n,RCS}}$), the number density of RCS segments longer than one Kuhn segment ($n_\text{RCS}$), and the orientational order parameter ($Q_{11}$). The results, normalized to their baseline values at $m^*$, are presented in Fig. \ref{fig:normalized-vs-m-star}.

As expected, a looser domain ($m = 1.1\, m^*$) increases the ACS mass fraction by capturing more atoms. This inclusion reduces the number average mass and number of segments in RCS.  Conversely, a tighter domain ($m = 0.9\, m^*$) has the opposite effect. Crucially, for the $\pm 10\%$ variation, all observed changes in these parameters remain within a $\pm 15\%$ window of the baseline. Even for the more extreme $\pm 20\%$ variation, the trends remain consistent and physically meaningful.

This analysis demonstrates that while the absolute values of the different  substructures have a modest dependence on the precise cutoff, their existence and the qualitative trends between them are an intrinsic characteristic of the system. Therefore, the classification into ACS, RCS, and CE is  not an artifact of a specific numerical choice but a genuine result of the underlying conformational heterogeneity within the melt.

Using this approach, we evaluated the data in Table~\ref{tab:acs-rcs-summary} and examined the orientational relaxation and translational dynamics described below. As a numerical observation, we note that the product $n_\text{RCS} k_\text{B}T$ has the dimensions of an energy density (\si{J\,m^{-3}} or Pa). For $T=600$~K, this gives 1.67~MPa. 
We do not interpret this quantity as a plateau modulus. However its numerical proximity to the reported plateau moduli for polydisperse polyethylene (1.00–2.58 MPa, Table 8 in Ref.~\onlinecite{Liu-Ruymbeke-2006}) is intriguing and may justify further investigation in future work.

\begin{table}[h!]
	\centering
	\caption{Mean and standard deviation values for structural parameters evaluated with the inner confinement domain definition.  Numerical precision is limited to the number of significant digits justified by the statistical variability of each quantity.}
	\begin{tabular}{l r r}
		\hline
		Parameter                          & Average      & STD \\
		\hline
		ACS mass fraction (\%)             & 36.14        & 0.02 \\
		$M_\text{n,RCS}$ (g\,mol$^{-1}$)   & 619          & 18 \\
		$M_\text{w,RCS}$ (g\,mol$^{-1}$)   & 731          & 33 \\
		$M_\text{z,RCS}$ (g\,mol$^{-1}$)   & 960          & 52 \\
		$n_{\text{RCS}}$ (nm$^{-3}$)       & 0.201        & 0.005  \\
		$Q_{11}$                           & 0.691        & 0.011 \\
		\hline
	\end{tabular}
	\label{tab:acs-rcs-summary}
\end{table}

\subsection{Local Conformational Dynamics and Torsional Cooperativity}

\paragraph*{Conformational Substructure and Heterogeneity.}
Although the Kuhn segment is the minimal statistical segment, it exhibits a broad distribution of conformational states. The $3^9$ possible torsional sequences for a segment of 11 bonds [Fig. \ref{fig:probability-blocks-11-to-110}(a)] give rise to a wide range of end-to-end distances, from compact, coiled structures ($\approx 4$ \AA) to highly extended ones ($\approx 14$ \AA). This intrinsic heterogeneity organizes into larger-scale substructures in the melt, specifically sequences of aligned segments (ACS) that aggregate into short-range ordered regions. \cite{Rigby-Roe-1995, Martins-macromol-2013, Kavassalis-Sundararajan-macrom-1993, Migler-trans-rich-2015}

\begin{figure}[h!]
	\centering
	\includegraphics[width=0.45\textwidth]{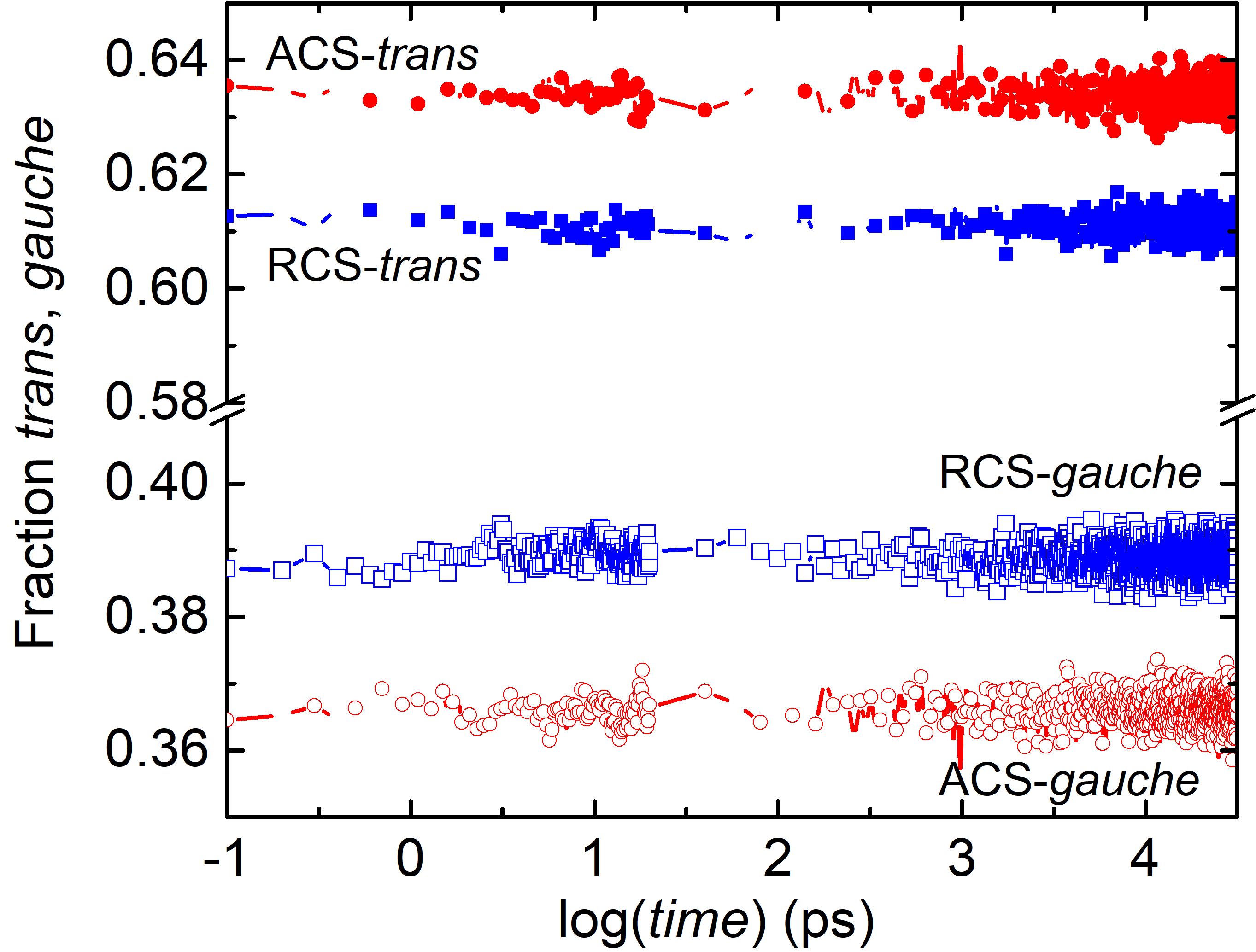}
	\caption{Fraction of the \textit{trans} and \textit{gauche} conformations in the ACS and RCS.}
	\label{fig:ACS-RCS-trans-gauche}
\end{figure}

\paragraph*{Dihedral Populations vs. Spatial Arrangement.}
Surprisingly, a quantitative analysis of dihedral states reveals that the distinction between ACS and RCS is not merely a difference in \textit{trans}/\textit{gauche} population. As shown in Fig. \ref{fig:ACS-RCS-trans-gauche}, the fraction of \textit{trans} conformations in ACS is only $\approx 3\%$ higher than in RCS. This residual difference is insufficient to explain their different orientational relaxation and translational dynamics illustrated below. It suggests that spatial arrangement and cooperativity of the dihedrals, rather than their relative populations, are the defining factors. 

\paragraph*{Impact on Orientational Dynamics.}
This conformational substructure has a profound impact on chain dynamics. The orientational relaxation for the different segments, characterized by the second-order Legendre function $C_2(t)$, is illustrated in Fig. \ref{fig:c2-c250-Kuhn-ACS-RCS-CE}. It clearly shows that the CE and RCS relax faster, and the ACS relax slowly. The relaxation of all these segments is not a Debye process and its detailed analysis is left for a future work. The fit of the ACS data with an exponential, $C_2(t) = A\, \exp(-t/\tau)$, is shown by the dotted line, yielding an average relaxation time $<\tau> = 30.7$ ps.

\subsection{Stretched Exponential Decay and Dynamic Heterogeneity}

\paragraph*{Stretched Exponential Fitting (KWW).}
An adequate description of the data required a stretched exponential (Kohlrausch-Williams-Watts) function:
\begin{equation}
		C_2(t) = A\, \exp\left[-\left(\frac{t}{\tau}\right)^\beta\right],
		\label{eq:stretched-exponential}
\end{equation}
where $\tau$ is a characteristic time scale, $\beta$ is the stretching exponent, and the pre-factor $A$ is set equal to one. The case $\beta = 1$ corresponds to single-exponential (Debye-type) relaxation, whereas $\beta < 1$ indicates a distribution of relaxation times. The mean relaxation time, calculated by integrating Eq. (\ref{eq:stretched-exponential}), is given by $ \langle \tau \rangle = \frac{\tau}{\beta} \, \Gamma(1/\beta) $.

\begin{figure}[h!]
	\centering
	\includegraphics[width=0.45\textwidth]{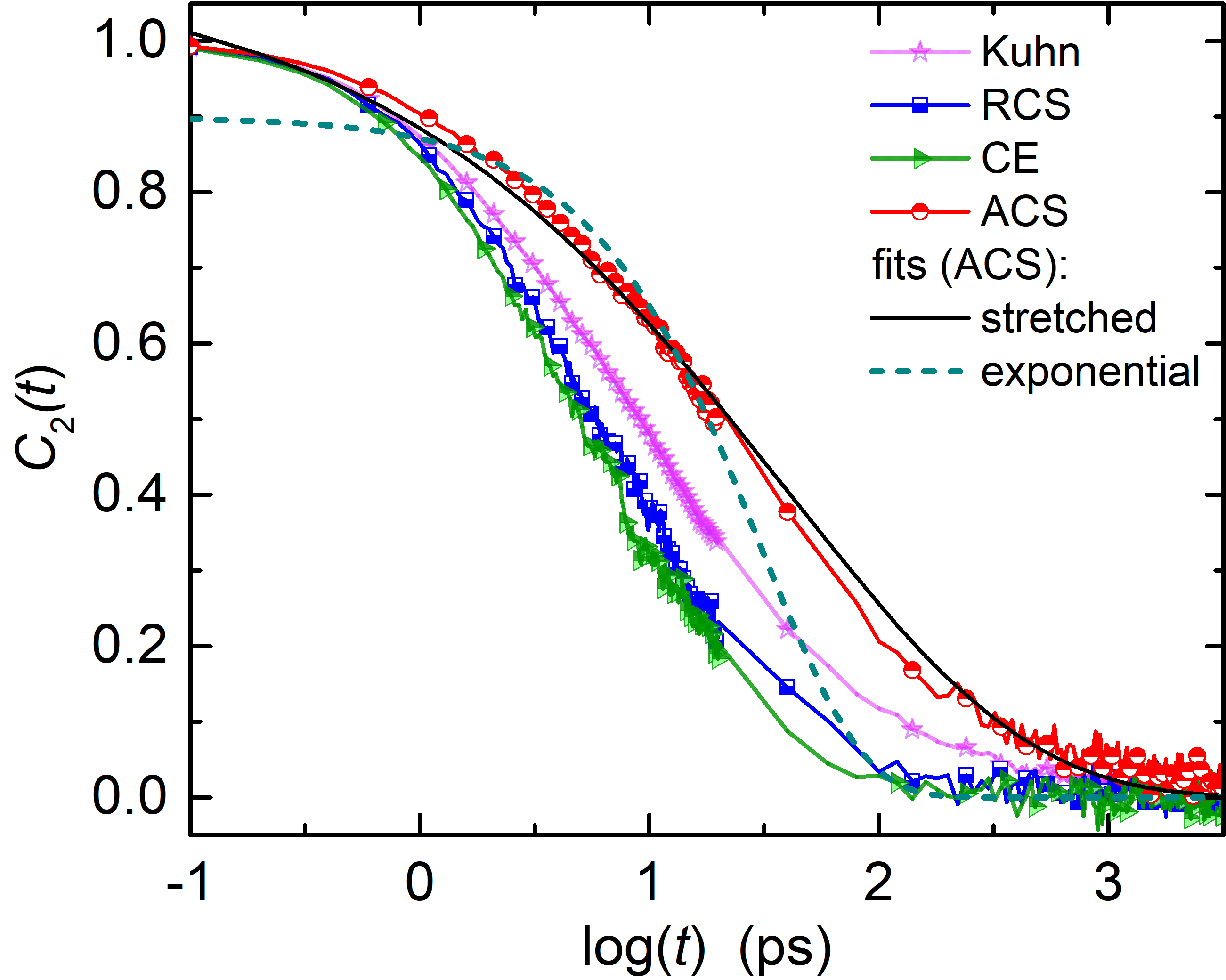}
	\caption{Orientational correlation function $ C_2(t) $ for ACS, RCS, and CE segments, illustrating the distinct relaxation dynamics. Only the fit to the ACS relaxation are illustrated. Fits for RCS and CE not shown due to visual clarity. }
	\label{fig:c2-c250-Kuhn-ACS-RCS-CE}
\end{figure}

The fits are shown as lines in Fig. \ref{fig:c2-c250-Kuhn-ACS-RCS-CE} (only for the ACS), and summarized in Table \ref{tab:relaxation-params}. They reveal a clear dynamic hierarchy. ACS segments relax the slowest $\langle \tau \rangle = 104.0$ ps, while the RCS and CE relax nearly an order of magnitude faster $\langle \tau \rangle = 13.2$ and 11.4 ps, respectively.

\begin{table}[h!]
	\centering
	\caption{Stretched exponential parameters and mean relaxation times for different types of Kuhn segments. The last row corresponds to the average results for all Kuhn segments. The criteria for numerical precision are the same as in Table \ref{tab:acs-rcs-summary}.  }
	\label{tab:relaxation-params}
	\begin{tabular}{lccc}
		\hline
		Segment Type & $ \tau $ (ps) & $ \beta $ & $ \langle \tau \rangle $ (ps) \\
		\hline
		ACS           & 48.8  & 0.484 & 104  \\
		RCS           & 10.5  & 0.700 & 13.2 \\
		CE            & 8.89  & 0.690 & 11.4 \\
		All Segments  & 16.8  & 0.613 & 24.1 \\
		\hline
\end{tabular}
\end{table}

\paragraph*{Non-Exponential Decay and Dynamic Heterogeneity.}
The non-exponential decay, $\beta < 1$, confirms that no single relaxation time can describe the dynamics; instead, there is a wide spectrum of local relaxation rates. Mathematically, $\exp \left[-\left(t/\tau_0\right)^\beta \right]$ is equivalent to a weighted average of simple exponential decays, where the weighting function is a broad distribution. The particularly low value of $\beta \approx 0.5$ for the ACS segments indicates an especially wide dispersion of rates, with a significant contribution from both very rapid and very slow relaxation processes.

\paragraph*{Nature of Dynamic Heterogeneity.}
Importantly, the stretched nature of the decay is not determined by the static, near-Gaussian block-scale distance statistics (see Section~\ref{sec:results-Gaussianity-of-blocks}). It is a manifestation of the dynamical heterogeneity that arises from different local conformational sequences within each segment type, ACS, RCS, and CE.

\subsection{Dihedral Flips and Localized Modes}

\paragraph*{Molecular Origin: The Localized-Mode Picture of Skolnick and Helfand.}
The molecular origin of this dynamical heterogeneity is explained by the \textit{localized-mode} picture of Skolnick and Helfand. \cite{Skolnick-Helfand-1980,Helfand-kinConfTrans-1971} A simplistic description of one of their main results is that a dihedral rotation is not an isolated event, but triggers correlated motions of neighboring bonds, and that the magnitude of these displacements decays rapidly with the distance from the transforming bond (see Table II in Refs.  \onlinecite{Skolnick-Helfand-1980,Helfand-kinConfTrans-1971}). The \textit{localized mode} is the reaction coordinate for the collective motion of the atoms adjacent to the rotating bond that changes from one conformational state to another. The motion is correlated and decays spatially. This decay provides the physical support for the definition of statistical, uncorrelated, Kuhn segments.

\paragraph*{Kramers' Theory and Fast/Slow Transitions.}
Skolnick and Helfand developed a multidimensional extension of Kramers' reaction rate theory for the rate at which a particle crosses a double-welled potential barrier, where the stable conformations appear as minima. The reaction coordinate is the one described above, the wells are two conformational states, and the activated complex is the saddle point of the reaction coordinate (Fig. 7 in Ref.~\onlinecite{Skolnick-Helfand-1980}). 

Within this framework they identified fast and slow transitions. The fast transitions occur when \textit{gauche} bonds are in even positions relative to the transforming bond (last row in Table \ref{tab:confseq-Skolnick}). Their explanation is the decrease in the displacement of tails units for a given rotation. On the other hand, when the \textit{gauche} bonds are odd neighbors, the local chain geometry hinders the formation of the localized mode. The tail becomes more rigid, opposing the correlated motion required for the transition, which decreases the relaxation rate (first three rows in Table \ref{tab:confseq-Skolnick}). We associate the stiff, slow relaxing segments to the ACS, and the flexible, fast relaxing segments to the RCS.

\begin{table}[h!]
	\centering
		\caption{Transition rates for selected conformational sequences, demonstrating the sensitivity to local environment. The results refer to a system of 12 carbon atoms, and 9 conformations (one Kuhn segment). The rotating bond is the \textit{trans} conformation at the center. This movement  is described by a localized mode (involving bond length and angle distortion, besides rotation) of the bond and adjacent segments. To maintain the chain stability, the torsion in bonds an even number way (which are parallel to the central bond) is greater than in the odd neighboring bonds. Data extracted from Table V in Skolnick and Helfand.~\cite{Skolnick-Helfand-1980}}
	\label{tab:confseq-Skolnick}
	\begin{tabular}{lc}
		\hline
		Conformational Sequence & Rate (ns$^{-1}$) \\
		\hline
		$(tg^{+}tg^{-})t(g^{+}tg^{+}t)$ & 1.00 \\
		$(tg^{-}tg^{-})t(g^{-}tg^{-}t)$ & 0.87 \\
		$(tg^{+}tg^{-})t(g^{-}tg^{+}t)$ & 1.11 \\
		$(g^{+}tg^{+}t)t(tg^{+}tg^{\pm})$ & 6.76 \\
		\hline
	\end{tabular}
\end{table}

\paragraph*{Interpretation of ACS/RCS Behavior.}
These results support our definition for the different types of Kuhn segments, explaining also the reason why the residual $\approx3\%$ difference in \textit{trans} population between the ACS and RCS cannot explain their different relaxation behavior. This behavior is determined by the sequence of the conformational states within each segment and by the cooperative mechanics involving rotations and bond deformations. Moreover, the broad distribution of relaxation times reflected in $\beta < 1$ mirrors the heterogeneity of local conformational environments along the chain.

\paragraph*{The Need for a Collective Relaxation Model.}
A key point emphasized by Skolnick and Helfand is that the microscopic rate constant for a single conformational transition is governed by local factors: the torsional barrier, local force constants, and the curvature of the reaction coordinate. The effect of slow, long-wavelength modes in the polymer is residual.\cite{Skolnick-Helfand-1980}
However, this picture of local jumps over an energy barrier, involving correlated motions and cooperativity of neighboring segments, contrasts with the non-exponential relaxation of collective relaxation times observed in bulk systems.

\subsection{Arrhenian Locally, and Non-Arrhenian Globally}

\paragraph*{Boyd's Contribution: Transitions vs. Relaxation.}
Boyd, Gee and Jin\cite{Boyd-JCP-1994} addressed this issue. They demonstrated that the conformational transition rates in bulk polyethylene remain Arrhenius (with an activation energy described by a single torsional barrier), while the relaxation times for the decay of the torsional angle autocorrelation function display non-Arrhenius temperature dependence and require stretched-exponential fits (Fig. 15 in Ref.~\onlinecite{Boyd-JCP-1994}). They concluded that ``even though  correlated, the transitions are not simultaneous with respect to barrier crossing''.

\paragraph*{Dynamic Heterogeneity in Bulk Systems.}
Their rationalization centered on the emergence of dynamic heterogeneity and may be summarized in the following points: (i) the complete decay of the torsional relaxation function implies that \emph{fairly long} chain segments lose spatial memory; (ii) this requires the relaxation and participation of all bonds in the system; (iii) as temperature decreases, the transitions become more and more unevenly distributed along the chain, inducing a heterogeneity in the spatial distribution of jumps; and (iv) intermolecular packing appears to have a subtle, but important, effect on conformational transitions.

\subsection{Continuous-Time Random Walks and Kuhn-Segment Heterogeneity}

\paragraph*{Shlesinger and Montroll Contribution: Bridging Local Jumps and Stretched Exponential Decay (CTRW).}
The connection between localized events and heterogeneities was formalized by
Shlesinger and Montroll. \cite{Shlesinger-Montroll-1984} They demonstrated that the ensemble of localized, correlated conformational transitions and the resulting heterogeneity in the spatial distribution of jumps, as rationalized by Boyd et al. for bulk systems, can be statistically described by a continuous-time random walk (CTRW), which alternates between steps and pauses (Fig. 8 in Ref.~\onlinecite{Berthier-RMPhys-2011}). 

They defined two distribution functions, one for the steps and another for the pauses.  Assuming for the pausing time distribution a form with a long tail used in the theory of charge transport in amorphous materials, $\psi(t) \propto t^{-1-\beta}$, they obtained the KWW stretched exponential decay function, Eq.~(\ref{eq:stretched-exponential}), observed for the collective relaxation. They established the validity of this equation for a general $0<\beta<1$ in three dimensions [Eq. (29) in Ref.~
\onlinecite{Shlesinger-Montroll-1984}]. 

In one dimension that equation reduces to
\begin{equation}
		C_\text{2,1D}(t) =  \exp\left[-\left(\frac{t}{\tau}\right)^{\beta/2}  \right] 
	\label{eq:stretched-exponential-1D}
\end{equation}
Thus, we conclude that the largest possible value of the Kohlrausch-Williams-Watts exponent occurs for quasi-one-dimensional defect diffusion, yielding $\beta_\text{1D} = 1/2$.\footnote{The quasi-one-dimensional defect diffusion is understood as a locally constrained motion along the aligned substructure.}
This value implies an extremely broad distribution of relaxation times, with significant contributions from very slow and very fast processes. Thus, the work of Shlesinger and Montroll relates the localized reaction-coordinate theory of Helfand and Skolnick, the understanding of dynamic heterogeneity by Boyd \textit{et al.}, and the stretched-exponential relaxation results presented in this work.

\paragraph*{The Special Meaning of $\mathrm{\beta\approx1/2}$.}
The value  $\beta \approx 0.5$ evaluated for the ACS segments indicates that the orientational relaxation of these aligned substructures is governed by quasi-one-dimensional, defect-mediated cooperative rearrangements, in full agreement with the Helfand--Skolnick localized-mode picture. 

In contrast, the larger $\beta \approx 0.7$ observed for RCS and CE suggests that the orientational relaxation of these segments proceeds through more accessible, effectively higher-dimensional pathways. The stretching exponent $\beta$ thus encodes not only the breadth of the relaxation spectrum but also the dimensionality of the underlying localized dynamics.

\subsection{Structural Origin of Stretched Exponential Relaxation}

It is instructive to contrast these results with those obtained from coarse-grained polymer models such as the bead--spring simulations of Vela and Simmons.\cite{Vela-Simmons-macom-2020} In that representation, the absence of torsional degrees of freedom prevents the formation of conformational substructures. The observed stretching in their study arises solely from generic dynamical heterogeneity and caging, without explicit torsional degrees of freedom. By contrast, in our atomistic representation each Kuhn segment samples a rich set of possible configurations ($3^9$ sequences for 11 bonds), which organizes into conformational substructures (ACS, RCS, CE) with distinct relaxation signatures. This highlights a crucial difference: while stretched exponential relaxation is a general feature of disordered dynamics, in polymers with internal torsions the stretching exponent $\beta$ acquires specific structural meaning, reflecting the spatial arrangement and cooperativity of dihedrals within statistical segments.

\subsection{Translational Dynamics}

\begin{figure}[h!]
	\centering
	\includegraphics[width=0.45\textwidth]{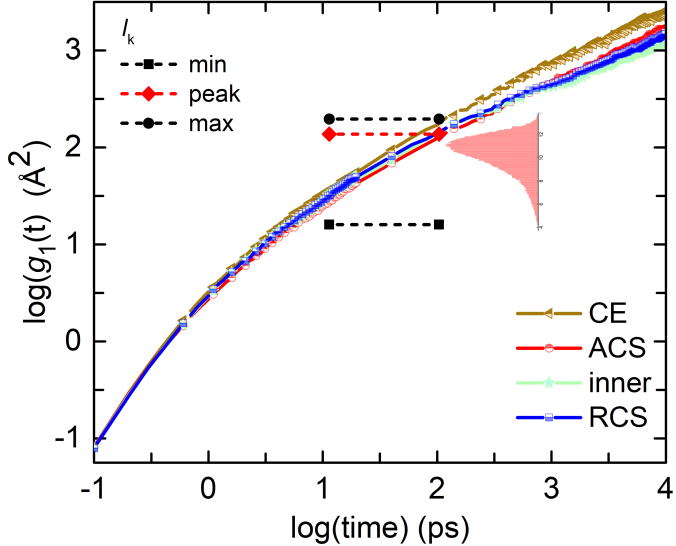}
	\caption{Mean-squared displacement of different types of Kuhn segments. Power-law slopes of $g_1(t)$ in the interval between the shortest (CE) and longest (ACS) orientational relaxation times are: CE = 0.751 ($R^2=0.991$), RCS = 0.688 ($R^2=0.990$), ACS = 0.715 ($R^2=0.984$). The RCS segments follows the dynamics of the inner chain  repeat units.  }
	\label{fig:g1-Kuhn-relax-time}
\end{figure}

The mean-squared displacement (MSD) of Kuhn segments, defined as $g_1(t) = \left\langle \left[ \mathbf{r}_i(t) - \mathbf{r}_i(0) \right]^2 \right\rangle$, 
where $\mathbf{r}_i$ is the center of mass of segment $i$, is shown in Fig. \ref{fig:g1-Kuhn-relax-time}. We focus on the time window spanned by the orientational relaxation times (Table \ref{tab:relaxation-params}) and on displacements comparable to the Kuhn segment size [Fig.~\ref{fig:probability-blocks-11-to-110}(a)]. Discussion of the results for other time windows is left for future work. 

\paragraph*{Dynamics of the ACS, RCS, and Middle-Chain Segments.}
At very short length scales (below $\approx 4$ \AA{}, the minimum Kuhn end-to-end distance), the motions of all segment types are indistinguishable. Differences in dynamical behavior emerge only when the segments start to explore displacements comparable to their own statistical size. In this key regime, chain ends (CE) diffuse most rapidly, followed by ACS, while RCS are the slowest. The effective slopes of $g_1(t)$ in this interval are subdiffusive and close to $0.7$. 

An important observation is that the dynamics of RCS segments and that of generic Kuhn segments at the middle of the chain are indistinguishable (the translational motion of the middle chain repeat units is used to identify the constrained regime).\cite{Kremer-Grest-1990,Ramos-2018,Harmandaris-2003} This agreement confirms that the  RCS segments represent the typical, bulk-like conformational state, exhibiting the random-walk statistics and subdiffusive dynamics characteristic of entropic springs in a dense melt.

\paragraph*{Relationship Between Subdiffusion and the Stretched Exponential.}

This analysis of translational dynamics reinforces the picture established from orientational relaxation, but it is crucial to distinguish between these two processes. The orientational relaxation, governed by local and cooperative conformational transitions, is slower for the dynamically constrained ACS segments ($\beta \approx 0.5$) and faster for the more flexible RCS and CE segments ($\beta \approx 0.7$), which share similar coiled conformations. 

The difference in their translational motion arises from their distinct positions within the chain. A flexible RCS segment is located between two slower, more rigid ACS. 
Although these ACS are not translationally immobile, their more cooperative, collective dynamics effectively constrain the motion of the adjacent RCS, temporarily limiting the available pathways for its center-of-mass diffusion. This mechanism plausibly explains why RCS segments, despite their faster orientational relaxation, exhibit the slowest translational diffusion. Furthermore, Fig. \ref{fig:g1-Kuhn-relax-time} shows that the $g_1(t)$ values for more coiled conformations are smaller than those for most extended ones. This behavior will be explored further in future works for larger mean-squared displacements and longer times. 

This interpretation is also supported by the conjecture of Boyd \textit{et al.}\cite{Boyd-JCP-1994}, who proposed that ``in the fast relaxing segments there is a decrease in the displacement of tail units for a given rotation''. In our case, this implies that RCS segments, being more coiled, move a shorter distance than ACS segments within the same time interval. The ACS, by contrast, are more aligned and can translate over longer distances, even though a small fraction of them participates in short-range ordered regions\cite{Martins-macromol-2013} and experiences enhanced friction due to van der Waals interactions.\cite{Martins-2011}

The observed subdiffusive behavior, with slopes near 0.7 for all segment types, is a hallmark of complex fluids reflecting transient trapping in heterogeneous environments.\cite{Berthier-RMPhys-2011,Ediger-annualreviews-2000} As reported by \citeauthor{Vidal-Russel-Israeloff-2000},\cite{Vidal-Russel-Israeloff-2000} the dynamics are intermittent in time, switching between moments of intense activity and moments of no dynamics at all. This observation suggests that ``extended regions of space transiently behave as fast and slow regions''.\cite{Berthier-RMPhys-2011} For polyvinylacetate, the size estimated for these regions was in the range between 2--3 nm, comprising cluster sizes of 30--90 repeat units.\cite{Vidal-Russel-Israeloff-2000} These values match the size of the short-range ordered regions previously identified, and also agree with the number of repeat units in the region, and a minimum of 75 repeat units for polyethlene (Figures 5b and 6b in Ref. \onlinecite{Martins-macromol-2013}). 

However, the relative ordering of the MSD curves (CE $>$ ACS $>$ RCS) reveals how the dynamic constraints imposed by local environment govern translational mobility. This interplay between segmental orientation, local environment, and center-of-mass motion establishes a direct connection between the stretched-exponential relaxation observed in orientation and the subdiffusive translational dynamics.

\paragraph*{Dynamical Heterogeneities at the Level of One Kuhn Segment.}
The above results (subdiffusive behavior and heterogeneous dynamics) can be viewed in the broader context in studies of glass-forming liquids, where particle displacements deviate from Gaussian statistics and mobility is spatially heterogeneous. \cite{Ediger-annualreviews-2000,Glotzer-PRL-1997,Glotzer-PR-E-2001, Berthier-RMPhys-2011} Glotzer and co-workers\cite{Glotzer-PR-E-2001} observed in simulations that mobile particles form transient stringlike clusters that lead to non-Fickian mean-squared displacements prior to the onset of long-time diffusion. Langer and Mukhopadhyay \cite{Langer-PhysRevE-2008} established a close link between this anomalous diffusion and the KWW function.

Our results demonstrate that an analogous heterogeneity exists intrinsically at the scale of a single Kuhn segment. However, while studies of glass-formers often reveal this heterogeneity phenomenologically, the present work identifies its specific microscopic origin in the conformational organization of the chain: ACS are orientationally rigid, while RCS and CE are more flexible. 

The distinct relaxation pathways are reflected in  their translational dynamics. The similar subdiffusive exponents for ACS and RCS suggest a complex coupling between orientational and center-of-mass motion. A compelling hypothesis, to be explored in future work, is that the ACS, with their quasi-one-dimensional relaxation, may translate via more concerted, string-like displacements. In contrast, the RCS, despite their faster orientational relaxation, may be translationally hindered by their coiled geometry and confinement between slow ACS. This picture suggests that the stretching exponent $\beta$ and the details of subdiffusion both encode the dimensionality and cooperativity of the underlying conformational dynamics.

\section{Conclusions}
\label{sec:conclusions}

We have analyzed and discussed in detail fundamental concepts in polymer science: the statistical segment and the Kuhn segment. The distribution of the distance between segments containing different numbers of C--C bonds in a flexible polyethylene chain was fitted to a Gaussian distribution, which was further analyzed using different tests involving the evaluation of various moments of the distribution and other quantitative measures of the deviation from Gaussianity. This allowed us, for the first time, to determine the minimal sizes associated with a statistical segment, an entropic spring, and a Gaussian segment directly from atomistic data. The conclusions were as follows:

\begin{itemize}
	\item The statistical, monomer-based segment $b$ used in equations for the plateau modulus, tube diameter and relaxation times,\cite{Larson-McLeish-2003} being a central variable in the theoretical models for the dynamics in polymer melts, is not statistical, nor is it Gaussian.
	\item The smallest statistical segment is the Kuhn segment, but this segment is not Gaussian. Thus, it cannot describe the entropic spring behavior.
	\item The smallest segment that describes an entropic spring with mild non-Gaussianity contains two Kuhn segments.
	\item The smallest segment whose internal distance distribution is well described by a Gaussian function contains five Kuhn segments.
	\item Segments containing ten or more Kuhn segments fall within the fully Gaussian regime according to all statistical diagnostics used (higher moments, $\alpha_2$, Q–Q plots, and histogram fits).
\end{itemize}

Once the minimal statistical segment was defined, we ran along a chain in half-Kuhn segment steps. For each segment, we recorded the position of all internal atoms relative to an axis joining the  first and last atoms of the segment. This procedure allowed us to directly evaluate the Kuhn diameter and conceive a conceptual confinement domain--a volume defined by this diameter that spatially encloses all segment atoms.  This provides a molecular-level definition of the Kuhn segment diameter. 

The spatial probability distribution of the atomic positions within this domain revealed three distinct regions of conformational preference:
\begin{itemize}
	\item An ``aligned core'', the high probability region, containing the more extended Kuhn segments.
	\item A ``semi-ordered shell'', an intermediate region that includes both extended and some transition configurations.
	\item The ``disordered periphery'', the low probability outermost region, containing segments in coiled conformations, with some atoms reaching the surface of the conceptual confinement domain. 
\end{itemize}

Based on the identification of the above three different regions, we implemented a quantitative procedure to classify segments, which involved two key steps:
\begin{itemize}
	\item Defining an inner-confinement domain based on the probability distribution.
	\item  Developing a reliability test to validate this definition.
\end{itemize}
The size of the inner-confinement domain was quantified by evaluating the cumulative probability for the distance distribution from $d=0$ up to a critical distance $d_\text{crit}$, where this probability equals 63.2\% $(1-e^{-1})$. 

The reliability test of this inner-domain criterion allowed us to validate the procedure for identifying the heterogeneous organization at the level of a single Kuhn segment. This methodology is robust and can be extended to other polymer systems beyond polyethylene.

Using this new probabilistic evaluation, supported by the reliability test, we isolate the segments residing in the statistically most confined region. We again traversed the chain segment by segment, and classified each Kuhn segment based on whether its atoms were predominantly located inside or outside the inner-domain.  We defined the former as Aligned Chain Segments (ACS). When the positions of the segments remained outside the inner-domain, they were classified as Random Conformational Sequences (RCS) or Chain Ends (CE). 

Specifically, the RCS are Kuhn segments outside the inner-confinement domain between two contiguous ACS. This classification reveals a Kuhn-scale structural heterogeneity not accessible in traditional analyses.

We then studied the orientational relaxation and translational diffusion of these different Kuhn segments. The orientational relaxation was fitted to a KWW function, and the stretching parameter and relaxation times were evaluated. The translational diffusion was analyzed from time zero up to the longest relaxation time evaluated for one Kuhn segment, within a length scale defined by the Kuhn segment's minimum and maximum size. The following conclusions were drawn:

\begin{itemize}
	\item The CE relax faster than the RCS, while the ACS relax more slowly than the other types of Kuhn segments. The orientational relaxation time for the ACS is one order of magnitude higher in comparison to the relaxation times of the other segments.
	\item For all segments, $\beta < 1$, with values around 0.7 for the RCS and CE, and around 0.5 for the ACS at 600~K.
	\item Segments with the same number of atoms relax differently depending on the position of \textit{gauche} conformations with respect to the rotating bond. Relaxation is faster when these conformations are located in \textit{even} positions with respect to the rotating bond. These correspond to the RCS and CE. In the ACS, the \textit{gauche} conformations are located at \textit{odd} positions. This interpretation is based on the results of Skolnick-Helfand\cite{Skolnick-Helfand-1980}. It explains why an increase of $\approx3\%$ in the fraction of \textit{trans} conformations in the ACS cannot explain the different relaxation of these segments. Skolnick and Helfand established that the a simple event of a dihedral flip is a localized collective excitation. This ``localized mode'' is a damped wave of atomic displacements that enables this transition to occur by minimizing the collective motion of the polymer chain. The damping of this wave defines the uncorrelated statistical segments, in particular the Kuhn segment. Boyd et al.\cite{Boyd-JCP-1994} demonstrated this excitation is locally Arrhenian and is described globally by a KWW function.
	\item Shlesinger and Montroll\cite{Shlesinger-Montroll-1984} demonstrated that the localized mode can be described by a continuous time random walk, with pulses and pauses, and that if the pausing distribution function has a long tail, then the decay function is the KWW stretched exponential. In one dimension, this implies that the maximum value of the stretched exponent is $\beta_{\text{1D}}=0.5$. This matches the value we obtained for the ACS at 600~K. This establishes a direct molecular interpretation for the stretched exponent in polymer melts.
\end{itemize}

Based on these results, we conclude that in melts of flexible polymer chains, the chain dynamics and segmental relaxation emerge from the interplay between localized excitations, conformational statistics, and the spatial organization of Kuhn segments. The present work provides the first molecular-level framework that unifies these features, revealing both the minimal statistical units of polymer models and the intrinsic heterogeneity at the Kuhn scale that governs relaxation in polymer melts.

\section*{Acknowledgements}
I acknowledge discussions with Loic Hillou and Sacha Mould.

\section*{AUTHOR DECLARATIONS}
\subsection*{Conflict of Interest}
The author have no conflicts to disclose.

\section*{Data Availability}
The data that support the findings of this study are available within the article and its supplementary material and are available from the author upon reasonable request.

\bibliography{references-KD.bib}

@article{Agapov-Sokolov-macrom-2010,
	author = {Agapov, A. and Sokolov, A. P.},
	title = {Size of the Dynamic Bead in Polymers},
	journal = {Macromolecules},
	volume = {43},
	number = {21},
	pages = {9126-9130},
	year = {2010},
	doi = {10.1021/ma101222y},
}

@article{Berthier-RMPhys-2011,
	title = {Theoretical perspective on the glass transition and amorphous materials},
	author = {Berthier, Ludovic and Biroli, Giulio},
	journal = {Rev. Mod. Phys.},
	volume = {83},
	issue = {2},
	pages = {587--645},
	numpages = {0},
	year = {2011},
	month = {Jun},
	publisher = {American Physical Society},
	doi = {10.1103/RevModPhys.83.587},
	url = {https://link.aps.org/doi/10.1103/RevModPhys.83.587}
}

@article{Boyd-JCP-1994,
	author = {Boyd, Richard H. and Gee, Richard H. and Han, Jie and Jin, Yong},
	title = {Conformational dynamics in bulk polyethylene: A molecular dynamics simulation study},
	journal = {The Journal of Chemical Physics},
	volume = {101},
	number = {1},
	pages = {788-797},
	year = {1994},
	month = {07},
	doi = {10.1063/1.468134},
	url = {https://doi.org/10.1063/1.468134},
}

@book{Dealy-Read-Larson-2018,
	address = {Cincinnati},
	edition = {2nd edition},
	title = {Structure and rheology of molten polymers: from structure to flow behavior and back again},
	isbn = {9781569906118},
	shorttitle = {Structure and rheology of molten polymers},
	publisher = {Hanser Publishers},
	author = {Dealy, John M. and Read, Daniel J. and Larson, Ronald G.},
	year = {2018},
}

@article{Ding-Sokolov-macrom-2004,
	author = {Ding, Yifu and Kisliuk, A. and Sokolov, A. P.},
	title = {When Does a Molecule Become a Polymer?},
	journal = {Macromolecules},
	volume = {37},
	number = {1},
	pages = {161-166},
	year = {2004},
	doi = {10.1021/ma035618i}
}

@article{Ding-Sokolov-JApplPolySci-2004,
	author = {Ding, Y. and Sokolov, A. P.},
	title = {Comment on the dynamic bead size and Kuhn segment length in polymers: Example of polystyrene},
	journal = {Journal of Polymer Science Part B: Polymer Physics},
	volume = {42},
	number = {18},
	pages = {3505-3511},
	doi = {10.1002/polb.20235},
	year = {2004}
}

@book{Doi-Edwards-ThePolymDyn-1994,
	author = {Doi, Masao and Edwards, Samuel F.},
	title = {The Theory of Polymer Dynamics},
	publisher = {Oxford Univ. Press Inc},
	year = {1994},
	address = {New York}
}

@article{Ediger-annualreviews-2000,
	title={Spatially Heterogeneous Dynamics in Supercooled Liquids},
	author={Ediger, MD},
	journal={Annual Review of Physical Chemistry},
	volume={51},
	number={1},
	pages={99-127},
	year={2000},
	doi={10.1146/annurev.physchem.51.1.99}
}

@article{Everears-Grest-R2/r-JCPhys-2003,
	author = {Auhl, Rolf and Everaers, Ralf and Grest, Gary S. and Kremer, Kurt and Plimpton, Steven J.},
	title = {Equilibration of long chain polymer melts in computer simulations},
	journal = {The Journal of Chemical Physics},
	volume = {119},
	number = {24},
	pages = {12718-12728},
	year = {2003},
	doi = {10.1063/1.1628670},
}

@article{Fatou-1965,
	author = {Fatou, J. G. and Mandelkern, L.},
	title = {The Effect of Molecular Weight on the Melting Temperature and Fusion of Polyethylene1},
	journal = {The Journal of Physical Chemistry},
	volume = {69},
	number = {2},
	pages = {417-428},
	year = {1965},
	doi = {10.1021/j100886a010}
	}

@incollection{Fetters-2007,
	author = {Fetters, L. J. and Lohsey, D. J. and Colby, R. H.},
	title = {Chain Dimensions and Entanglement Spacings},
	editor = {Mark, J.},
	booktitle = {Physical Properties of Polymers Handbook},
	edition = {2nd},
	publisher = {Springer},
	chapter = {25},
	year = {2007}
}

@book{Flory-PPChem-1953, 
	place={Ithaca, New York}, 
	title={Principles of polymer chemistry}, 
	publisher={Cornell University Press}, 
	author={Flory, Paul J.}, 
	year={1953}
}

@article{Flory-JCP-1949,
	author = {Flory, Paul J.},
	title = {The Configuration of Real Polymer Chains},
	journal = {The Journal of Chemical Physics},
	volume = {17},
	number = {3},
	pages = {303-310},
	year = {1949},
	month = {03},
	doi = {10.1063/1.1747243},
}

@book{Gedde-2019, 
	author = {Gedde, U.W. and Hedenqvist, M.S.},
	title = {Fundamental Polymer Science}, 
	edition ={2nd},
	year = {2019}, 
	publisher = {Springer-Nature Switzerland AG}
}

@article{Glotzer-PRL-1997,
	title = {Dynamical Heterogeneities in a Supercooled Lennard-Jones Liquid},
	author = {Kob, Walter and Donati, Claudio and Plimpton, Steven J. and Poole, Peter H. and Glotzer, Sharon C.},
	journal = {Phys. Rev. Lett.},
	volume = {79},
	issue = {15},
	pages = {2827--2830},
	numpages = {0},
	year = {1997},
	month = {Oct},
	publisher = {American Physical Society},
	doi = {10.1103/PhysRevLett.79.2827},
	url = {https://link.aps.org/doi/10.1103/PhysRevLett.79.2827}
}

@article{Glotzer-PR-E-2001,
	title = {Spatially correlated dynamics in a simulated glass-forming polymer melt: Analysis of clustering phenomena},
	author = {Gebremichael, Y. and Schr\o{}der, T. B. and Starr, F. W. and Glotzer, S. C.},
	journal = {Phys. Rev. E},
	volume = {64},
	issue = {5},
	pages = {051503},
	numpages = {13},
	year = {2001},
	month = {Oct},
	publisher = {American Physical Society},
	doi = {10.1103/PhysRevE.64.051503},
	url = {https://link.aps.org/doi/10.1103/PhysRevE.64.051503}
}

@article{Harmandaris-2003,
	author = {Harmandaris, V.A. and Mavrantzas, V.G. and Theodorou, D.N. and Kröger, M. and Ramírez, J. and Öttinger, H.C. and Vlassopoulos, D.},
	title = {Crossover from the Rouse to the entangled polymer melt regime: signals from long, detailed atomistic molecular dynamics simulations, supported by rheological experiments},
	year = {2003},
	volume = {36},
	pages = {1376–1387},
	journal = {Macromolecules}
}

@article{Helfand-kinConfTrans-1971,
	author = {Helfand, E.},
	title = {Theory of the kinetics of conformational transitions in polymers},
	year = {1971},
	volume = {54},
	pages = {4651–4661},
	journal = {The Journal of Chemical Physics},
	doi = {10.1063/1.1674737,}
}

@article{Hess-2008a,
	author = {Hess, B. and Kutzner, C. and Spoel, D. and Lindahl, E.},
	title = {GROMACS 4: algorithms for highly efficient, load-balanced, and scalable molecular simulation},
	year = {2008},
	volume = {4},
	pages = {435–447},
	journal = {J. Chem. Theory Comput}
}

@article{Hsu-Kremer-2016,
	author = {Hsu, Hsiao-Ping and Kremer, Kurt},
	title = {Static and dynamic properties of large polymer melts in equilibrium},
	journal = {The Journal of Chemical Physics},
	volume = {144},
	number = {15},
	pages = {154907},
	year = {2016},
	month = {04},
	issn = {0021-9606},
	doi = {10.1063/1.4946033}
}

@article{Ionue-macrom-2002,
	author = {Inoue, Tadashi and Uematsu, Takehiko and Osaki, Kunihiro},
	title = {The Significance of the Rouse Segment: Its Concentration Dependence},
	journal = {Macromolecules},
	volume = {35},
	number = {3},
	pages = {820-826},
	year = {2002},
	doi = {10.1021/ma011037m}
}

@article{Jayaraman-macrom-2019,
	author = {Gartner, Thomas
	E. III and Jayaraman, Arthi},
	title = {Modeling and Simulations of Polymers: A Roadmap},
	journal = {Macromolecules},
	volume = {52},
	number = {3},
	pages = {755-786},
	year = {2019},
	doi = {10.1021/acs.macromol.8b01836}	
}

@article{Langer-PhysRevE-2008,
	title = {Anomalous diffusion and stretched exponentials in heterogeneous glass-forming liquids: Low-temperature behavior},
	author = {Langer, J. S. and Mukhopadhyay, Swagatam},
	journal = {Phys. Rev. E},
	volume = {77},
	issue = {6},
	pages = {061505},
	numpages = {8},
	year = {2008},
	month = {Jun},
	publisher = {American Physical Society},
	doi = {10.1103/PhysRevE.77.061505},
}

@article{Kavassalis-Sundararajan-macrom-1993,
	author = {Kavassalis, T. A. and Sundararajan, P. R.},
	title = {A molecular-dynamics study of polyethylene crystallization},
	journal = {Macromolecules},
	volume = {26},
	number = {16},
	pages = {4144-4150},
	year = {1993},
	doi = {10.1021/ma00068a012},
}

@article{Kremer-Grest-1990,
	author = {Kremer, K. and Grest, G.S.},
	title = {Dynamics of linear polymer melts: a molecular dynamics simulation},
	year = {1990},
	volume = {92},
	pages = {5057–5086},
	journal = {The Journal of Chemical Physics},
	doi ={10.1063/1.458541},
}

@article{Kruteva-Zamponi-macom-2021,
	author = {Kruteva, Margarita and Zamponi, Michaela and Hoffmann, Ingo and Allgaier, J{\"u}rgen and Monkenbusch, Michael and Richter, Dieter},
	title = {Non-Gaussian and Cooperative Dynamics of Entanglement Strands in Polymer Melts},
	journal = {Macromolecules},
	volume = {54},
	number = {24},
	pages = {11384-11391},
	year = {2021},
	doi = {10.1021/acs.macromol.1c01859},
}

@article{Larson-McLeish-2003,
	author = {Larson, R. G. and Sridhar, T. and Leal, L. G. and McKinley, G. H. and Likhtman, A. E. and McLeish, T. C. B.},
	copyright = {The Society of Rheology},
	issn = {0148-6055},
	journal = {Journal of Rheology (New York : 1978)},
	pages = {809-818},
	title = {Definitions of entanglement spacing and time constants in the tube model},
	volume = {47},
	year = {2003},
	number = {3},
	doi ={10.1122/1.1567750},
}

@article{Likhtman-2002,
	author = {Likhtman, Alexei E. and McLeish, Tom C. B.},
	title = {Quantitative Theory for Linear Dynamics of Linear Entangled Polymers},
	journal = {Macromolecules},
	volume = {35},
	number = {16},
	pages = {6332-6343},
	year = {2002},
	doi = {10.1021/ma0200219},
	URL = { 
	https://doi.org/10.1021/ma0200219
	}	
}

@article{Liu-Ruymbeke-2006,
	title = {Evaluation of different methods for the determination of the plateau modulus and the entanglement molecular weight},
	journal = {Polymer},
	volume = {47},
	number = {13},
	pages = {4461-4479},
	year = {2006},
	issn = {0032-3861},
	doi = {https://doi.org/10.1016/j.polymer.2006.04.054},
	url = {https://www.sciencedirect.com/science/article/pii/S0032386106005684},
	author = {Chenyang Liu and Jiasong He and Evelyne van Ruymbeke and Roland Keunings and Christian Bailly}
}

@book{Lodge-Heimenz-PolyChem-2007,
	author = {Lodge, Timothy and Hiemenz, Paul C.},
	address = {Boca Raton, FL},
	edition = {Second edition.},
	publisher = {CRC Press, Taylor \& Francis Group},
	title = {Polymer Chemistry},
	year = {2007}
}

@article{Martins-2011,
	author = {José A. Martins},
	title = {Toward a Physical Definition of Entanglements},
	journal = {Journal of Macromolecular Science, Part B},
	volume = {50},
	number = {4},
	pages = {769--794},
	year = {2011},
	publisher = {Taylor \& Francis},
	doi = {10.1080/00222341003785151},	
}

@article{Martins-macromol-2013,
	author = {Martins, Jos{\'e} A. and Micaelo, Nuno M.},
	title = {Short-Range Order in Polyethylene Melts: Identification and Characterization},
	journal = {Macromolecules},
	volume = {46},
	number = {19},
	pages = {7977-7988},
	year = {2013},
	doi = {10.1021/ma4009934}	
}

@article{Migler-trans-rich-2015,
	author = {Migler, Kalman B. and Kotula, Anthony P. and Hight Walker, Angela R.},
	title = {Trans-Rich Structures in Early Stage Crystallization of Polyethylene},
	journal = {Macromolecules},
	volume = {48},
	number = {13},
	pages = {4555-4561},
	year = {2015},
	doi = {10.1021/ma5025895},
}

@article{Paul-1995a,
	author = {Paul, W. and Yoon, D.Y. and Smith, G.D.},
	title = {An optimized united atom model for simulations of polymethylene melts},
	year = {1995},
	volume = {103},
	pages = {1702–1709},
	journal = {J. Chem. Phys},
	doi ={10.1063/1.469740},
}

@article{Paul-1997a,
	author = {Paul, W. and Smith, G.D. and Yoon, D.Y.},
	title = {Static and dynamic properties of a n-C100H202 melt from molecular dynamics simulations},
	year = {1997},
	volume = {30},
	pages = {7772–7780},
	journal = {Macromolecules},
	doi = {10.1021/ma971184d},
}

@article{Ramos-2018,
	author = {J. Ramos and J.F. Vega and J. Martínez-Salazar},
	title = {Predicting experimental results for polyethylene by computer simulation},
	journal = {European Polymer Journal},
	volume = {99},
	pages = {298-331},
	year = {2018},
	issn = {0014-3057},
	doi = {10.1016/j.eurpolymj.2017.12.027},
}

@book{Riso-Statistical-wR-2008,
	author = {Rizzo, Maria L.},
	address = {Boca Raton},
	booktitle = {Statistical computing with R},
	isbn = {9781584885450},
	publisher = {Chapman \& Hall/CRC},
	series = {Chapman \& Hall/CRC computer science and data analysis series},
	title = {Statistical computing with R  / Maria L. Rizzo.},
	url = {http://catdir.loc.gov/catdir/enhancements/fy0740/2007034218-d.html},
	edition = {2nd edition},
	year = {2019},
}

@book{Rubinstein-Colby-PolPhys-2003,
	author = {Rubinstein, Michael and Colby, Ralph H.},
	title = {Polymer Physics},
	publisher = {Oxford University Press Inc},
	year = {2003},
	address = {New York}
}

@article{Vidal-Russel-Israeloff-2000,
    author = {Vidal Russell, E and Israeloff, E. N. },
    title ={Direct observation of molecular cooperativity near the glass transition},
    journal = { Nature}, 
    volume = {408}, 
    pages = {695–698}, 
    year =  {2000}, 
    doi = {10.1038/35047037},
}

@article{Shlesinger-Montroll-1984,
	author = {Michael F. Shlesinger  and Elliott W. Montroll},
	title = {On the Williams—Watts function of dielectric relaxation},
	journal = {Proceedings of the National Academy of Sciences},
	volume = {81},
	number = {4},
	pages = {1280-1283},
	year = {1984},
	doi = {10.1073/pnas.81.4.1280},
	}

@book{Sivia-Skilling-2006,
	address = {Oxford},
	edition = {2nd edition},
	isbn = {9780198568315},
	publisher = {Oxford University Press},
	series = {Oxford science publications},
	title = {Data analysis: a Bayesian tutorial.},
	url = {http://catdir.loc.gov/catdir/enhancements/fy0726/2006284782-b.html},
	year = {2006},
	author = {Sivia, D. S. and Skilling, J.},
}

@book{Stanford-QQ-plots-1994,
	author = {Stanford, John L. and Vardeman, Stephen B.},
	address = {San Diego},
	booktitle = {Statistical methods for physical science},
	publisher = {Academic Press},
	series = {Methods of experimental physics ; v. 28},
	title = {Statistical methods for physical science  / edited by John L. Stanford and Stephen B. Vardeman.},
	year = {1994}
}

@article{Skolnick-Helfand-1980,
	author = {Skolnick, J. and Helfand, E.},
	title = {Kinetics of conformational transitions in chain molecules},
	year = {1980},
	volume = {72},
	pages = {5489–5500},
	journal = {The Journal of Chemical Physics},
	doi = {10.1063/1.438965},
}

@Article{Sundararajan-Kavassalis-JCSoc-1995,
	author ={Sundararajan, P. R. and Kavassalis, T. A.},
	title  ={Molecular dynamics study of polyethylene chain folding: the effects of chain length and the torsional barrier},
	journal  ={J. Chem. Soc., Faraday Trans.},
	year  ={1995},
	volume  ={91},
	issue  ={16},
	pages  ={2541-2549},
	publisher  ={The Royal Society of Chemistry},
	doi = {10.1039/FT9959102541},
	}

@article{Rigby-Roe-1995,
	author = {Rigby, David and Roe, Ryong‐Joon},
	title = {Molecular dynamics simulation of polymer liquid and glass. II. Short range order and orientation correlation},
	journal = {The Journal of Chemical Physics},
	volume = {89},
	number = {8},
	pages = {5280-5290},
	year = {1988},
	month = {10},
	doi = {10.1063/1.455619},
}

@book{Teraoka-PolymSol-2002,
	author = {Teraoka, Iwao},
	address = {New York},
	publisher = {Wiley-Interscience},
	title = {Polymer solutions: an introduction to physical properties},
	year = {2002},
}

@book{Yamakawa-1971,
	author = {Yamakawa, Hiromi},
	address = {New York},
	isbn = {0060473096},
	publisher = {Harper \& Row},
	series = {Harper's chemistry series},
	title = {Modern theory of polymer solutions},
	year = {1971}
}

@article{Vela-Simmons-macom-2020,
	author = {Diaz Vela, Daniel and Simmons, David S.},
	title = {The microscopic origins of stretched exponential relaxation in two model glass-forming liquids as probed by simulations in the isoconfigurational ensemble},
	journal = {The Journal of Chemical Physics},
	volume = {153},
	number = {23},
	pages = {234503},
	year = {2020},
	doi = {10.1063/5.0035609},
}

@article{Wischnewski-2003a,
	author = {Wischnewski, A. and Monkenbusch, M. and Willner, L. and Richter, D.},
	title = {Direct observation of the transition from free to constrained single-segment motion in entangled polymer melts},
	year = {2003},
	volume = {90},
	pages = {58302},
	journal = {Phys. Rev. Lett}
}

\end{document}